\documentclass[fleqn,usenatbib]{mnras}

\usepackage{newtxtext,newtxmath}
\usepackage[T1]{fontenc}

\DeclareRobustCommand{\VAN}[3]{#2}
\let\VANthebibliography\thebibliography
\def\thebibliography{\DeclareRobustCommand{\VAN}[3]{##3}\VANthebibliography}

\usepackage{subfig}
\usepackage{graphicx}	% Including figure files
\usepackage{amsmath}	% Advanced maths commands
\usepackage{newtxtext,newtxmath}	% Extra maths symbols
\usepackage{multirow}
\usepackage{tabularx}
\usepackage{arydshln}

\title[Ejecting-crust activity model on comet 67P]{Constraints on the ejecting-crust activity model on comet 67P/Churyumov-Gerasimenko}

\author[N. Attree et al.]{
N. Attree$^{1}$\thanks{E-mail: attree@iaa.es},
P. Gutierrez$^{1}$,
C. Schuckart$^{2}$,
J. Markkanen$^{2}$,
Y. Skorov$^{2,4}$,
Y. Xin$^{5,6}$,
\newauthor
D. Bischoff$^{2}$,
B. Gundlach$^{3}$,
J. Blum$^{2}$
\\
$^{1}$Instituto de Astrofísica de Andalucía (CSIC), Glorieta de la
Astronomía s/n. 18008 Granada, Spain\\
$^{2}$Institut f\"ur Geophysik und Extraterrestrische Physik, Technische Universit\"at Braunschweig, Mendelssohnstr. 3, 38106 Braunschweig, Germany\\
$^{3}$Institut f\"ur Planetologie, Universit\"at M\"unster, Wilhelm-Klemm-Str. 10, 48149, M\"unster, Germany\\
$^{4}$Max-Planck-Institut f\"ur Sonnensystemforschung, Justus-von-Liebig-Weg 3, 37077, G\"ottingen, Germany\\
$^{5}$Key Laboratory of Planetary Sciences, Purple Mountain Observatory, Chinese Academy of Sciences, Nanjing, PR China\\
$^{6}$School of Astronomy and Space Science, University of Science and
Technology of China, Hefei 230026, PR China
}

\date{Accepted XXX. Received YYY; in original form ZZZ}

\pubyear{2024}

\begin{document}
\label{firstpage}
\pagerange{\pageref{firstpage}--\pageref{lastpage}}
\maketitle

\begin{abstract}
Reproducing the observed activity of comets with thermophysical models remains a primary challenge of cometary science. We use a pebble-based thermophysical model of gas-pressure build-up in the subsurface to reproduce the global emission rates of dust, water, CO$_{2}$, and CO observed by Rosetta at comet 67P/Churyumov-Gerasimenko (hereafter 67P). For sufficiently low diffusivities, the low tensile strength is overcome, leading to the ejection of $\sim$ millimetre- to decimetre-sized dust-particles as well as roughly the correct outgassing rates. All the ejections, and thus the bulk of the outgassing, come from the southern hemisphere during the time that it is strongly illuminated at perihelion. This leads to a 'blow-off' of the dust-crust that otherwise forms: volatiles are much closer to the surface in the south (within the top centimetre) than in the north (10-or-more cm deep), naturally explaining the strong southern water-outgassing expected from 67P's non-gravitational accelerations and torques. We find that low gas-diffusivity, as well as large heat-capacity and steeply decreasing tensile strength with depth or ice-content, are in best agreement with the outgassing data. However, even in these cases, we struggle not to exceed the observed emission rates of dust, CO$_{2}$, and CO. In the south, it is difficult for models to achieve a balance between triggering activity and generating too much of it (with CO$_{2}$ the critical driving-species here); while in the north, it remains challenging to generate activity at all. Strong constraints are placed on the nature of the activity mechanism by the location of dust-ejection and erosion. 
\end{abstract}

%%%%%%%%%%%%%%%%%%%%%%%%%%%%%%%%%%%%%%%%%%%%%%%%%%

%%%%%%%%%%%%%%%%% BODY OF PAPER %%%%%%%%%%%%%%%%%%

\section{Introduction}
\label{sec:introduction}

Cometary activity is an interesting and still not well-explained solar system phenomenon that is vital for understanding the composition, structure, and evolution of these relatively pristine bodies. As a comet nucleus is heated by the Sun, volatile ices sublimate, driving the ejection of refractory dust-particles that form the observable coma and tail. Thermophysical modelling has shown that gas sublimation pressures can exceed gravity on typical cometary nuclei, but that the material tensile strength or cohesion is much more difficult to overcome. This is the so-called cohesion bottleneck or activity paradox \citep{Kuehrt1994, Jewitt2019}, whereby dust ejection appears impossible and outgassing should gradually decline as a thick dust-crust builds up. Resolving this paradox is necessary for deciphering comet structure, composition, and evolution.

One potential solution to the cohesion bottleneck is the arrangement of the refractory and icy grains making up comets into millimetre- to centimetre- sized pebbles \citep{Blum2017} which, due to their size and number of inter-particle connections, have a low tensile strength \citep{Skorov.2012}. The $\sim15-30$ au region of the Solar System was rich in $\mu$m-sized ice grains, organics, silicates, sulphides, and metal, that coagulated to form cm-sized porous pebbles. \citet{Blum2006, Blum2008, Zsom2010} and \citet{Davidsson2016} outlined a comet formation scenario that starts in the solar nebula and ends in the primordial disk, that reproduces the observed properties, and additionally explains the presence of extensive layering on 67P/Churyumov-Gerasimenko (and on 9P/Tempel 1 observed by Deep Impact).

\citet{Gundlach.2020} modelled pressure build-up in a pebble-structured comet, as applied to 67P, the best studied comet due to the data gathered by the ESA Rosetta space mission. The pebble model was used to simulate 67P under conditions of constant, intense illumination relevant for its southern summer at perihelion. Continuous illumination indeed led to pressure build-up that was sufficient to overcome the low tensile strength and surface gravity, leading to the ejection of cm-and-above sized pebbles and chunks. Outgassing rates of water and CO$_{2}$ that roughly matched the Rosetta observations were also found. Subsequently, the model was extended to the differing illumination conditions of a full cometary orbit, but still for a single one-square metre surface patch \citep{bischoff2023}. With the patch located at the equator, it was found to be very difficult to generate enough pressure to overcome the tensile strength; conditions of long, slow insolation increase drove the volatiles deeper and deeper under the surface, without ever reaching the requisite pressures. When the tensile strength was set to very low values, or ejections were triggered by various artificial criteria, very high dust and CO$_{2}$ ejection rates (in some cases exceeding the water outgassing rate by many times) were found, again at odds with observations.

The model was further developed in \citet{Attree2024b}, in order to simulate pressure build-up inside the pebbles themselves, as proposed by \citet{Fulle2019} and \citet{Fulle2020}. The low diffusivity inside pebbles combined with their low strength (due to their small constituent grains but with low connectivity) allows high pressures to break the pebbles and erode the surface by the ejection of sub-pebble particles (in the size range of microns to millimetres), overcoming the cohesion bottleneck. While this mechanism was shown to be feasible, it does not explain the bulk activity mechanism on 67P's surface, the majority of which is not dominated by the Water Enriched Bodies (WEBs) where this erosion can take place \citep{Ciarniello2022, Ciarniello2023}. When the CO$_{2}$-rich, non-WEB material which makes up $\sim90\%$ of the surface in this model was simulated, chunk ejection was limited and CO$_{2}$ outgassing again exceeded that of water. An explanation for the non-WEB-driven bulk activity is therefore still required.

Despite the above difficulties, \citet{Attree2024b} did have some success generating ejections when using a low gas-diffusivity value, demonstrating the importance of this parameter. Likewise, it was shown that simulating multiple latitudes across the comet's surface was necessary in order to capture the very different illumination trends caused by 67P's strong seasons. Thus, here we continue the application of the same thermophysical model, concentrating on pressure in-between pebbles, and on results with the lower gas diffusivity value and at multiple points on 67P's surface. The model is also extended to include another volatile: CO, which we include as a separate ice species here. We do not focus on the internal pebble pressure here, as this was addressed in \citet{Attree2024b}.

The remainder of the paper is organised as follows: in section \ref{sec:TPM}, we give an overview of the thermophysical model, before describing the results of its application to 67P in section \ref{sec:results}. The interpretation of these results is discussed in section \ref{sec:discussion}, and conclusions drawn in section \ref{sec:conclusions}.

\section{Thermophysical model}
\label{sec:TPM}

The numerical model is described in \citet{Gundlach.2020}, \citet{bischoff2023}, and \citet{Attree2024b}. It solves the heat-transfer equation for temperature $T$ at time $t$ and depth $z$
\begin{equation}
    \rho(z) c(z) \frac{dT(z,t)}{dt} = \frac{d}{dz} \left[ \lambda(T(z),z) \frac{dT(z,t)}{dz} \right] + Q(T(z),z)
    \label{eq:heat_transport_eq}
\end{equation}
using the finite difference method and the forward difference scheme with constant depth- and time-steps, $dt$ and $dz$. In the nominal model runs, $dt=10$ s and the depth-step is set equal to the pebble radius, $dz=R=5$ mm, which complies with the convergence criterion for explicit methods \citep{Patankar}. We tested a run with $dt=1$ s and, as expected, found no significant differences in results. The numerical resolution of the depth-step is further discussed in Section \ref{sec:results:resolution}. The density, $\rho$, heat-capacity, $c$, and thermal conductivity, $\lambda$, of each numerical layer are determined by the local, time-varying content of ices and refractory dust. Initial abundances are defined in terms of a bulk density of $532 \,\mathrm{kg/m^3}$ and various values of dust-to-total ice mass-ratio, $\delta$, and CO$_2$-to-H$_{2}$O and CO-to-H$_{2}$O ice mass fractions, $f_{CO_2}$ and $f_{CO}$. The material is assumed to be made of pebbles of radius $R$, with internal volume filling factors of 0.4 (porosity 0.6) packed together with a volume filling factor of 0.6 (porosity 0.4), giving a total bulk porosity equal to 67P's $\sim76\%$. Thermal conductivity, $\lambda(T(z),z)$ in Eqn.~\ref{eq:heat_transport_eq}, is temperature dependent and given by the sum of the conductivity from connections between contacting pebbles (which depends on the volume filling factor/porosity) and the radiative component between them:
\begin{equation}
    \lambda = \lambda_{\mathrm{net}}(R) + \lambda_{\mathrm{rad}}(R,T) = \lambda_{\mathrm{net}}(R) + \lambda_{\mathrm{rad}}(R)\, \left( \frac{T}{\mathrm{1~K}} \right)^3,
    \label{eq:pebble_conductivity}
\end{equation}
where the parameters $\lambda_{\mathrm{net}}(R)$ and $\lambda_{\mathrm{rad}}(R)$ are described by the equations in \citet{Gundlach.2020}: section A1.1. The radiative term $\lambda_{\mathrm{rad}}$ dominates over the network conductivity for $\sim$mm-and-above-sized pebbles over $\sim100$ K.

\begin{figure}
\resizebox{\hsize}{!}{\includegraphics{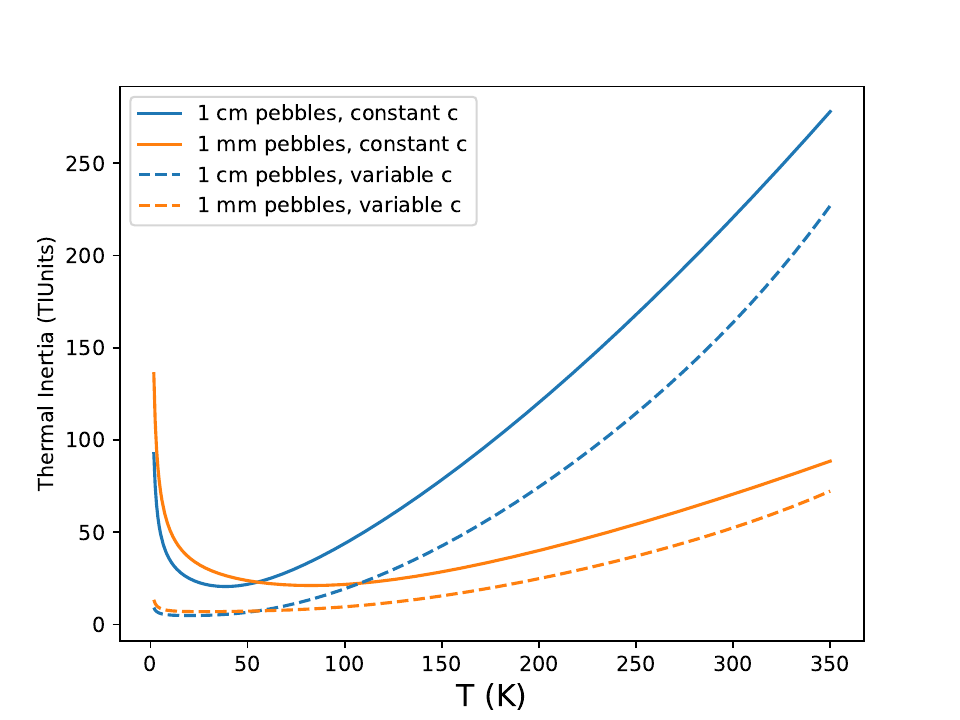}}
\caption{Thermal inertia, in SI units of Jm$^{-2}$s$^{-0.5}$K$^{-1}$, for two different pebble sizes with fixed $c_{dust}=3000$ J kg$^{-1}$ K$^{-1}$, and the temperature-dependent heat capacities from \citet{bischoff2023}.}
\label{plot_thermalinertia}
\end{figure}

In the previous paper \citep{Attree2024b}, temperature-independent heat capacities were assumed for each component, respectively a fixed $c_{dust}=3000$ J kg$^{-1}$ K$^{-1}$,  $c_{H_{2}O}=1610$ J kg$^{-1}$ K$^{-1}$, and $c_{CO_{2}}=850$ J kg$^{-1}$ K$^{-1}$. In \citet{bischoff2023}, however, temperature dependent values were used. While the latter two fixed values approximate well the temperature-dependent curves used in \citet{bischoff2023}, the former value for dust is rather high (as compared to those measured for meteorites for example, see references in \citealp{bischoff2023}). We investigate this briefly here by plotting the thermal inertia, $I=\sqrt{c\rho\lambda}$ for two different pebble sizes, using $c_{dust}=3000$ J kg$^{-1}$ K$^{-1}$ and the variable $c_{H_{2}O}$, $c_{CO_{2}}$, and $c_{dust}$ from equations A1-3 of \citet{bischoff2023}, and a dust-to-total-ice mass ratio $\delta=2$, in Figure \ref{plot_thermalinertia}. The high temperature-dependence of the thermal conductivity given by Eqn.~\ref{eq:pebble_conductivity} dominates over the roughly linear temperature dependencies of the heat capacities, as can be seen by the similar gradients between the constant (solid lines) and variable (dashed lines) curves as temperature increases, so that the temperature dependence of the latter has little effect. The lower overall values in the variable heat-capacity model, however, deserve some attention. For comparison, a wide range of thermal inertias  between zero and 350 Jm$^{-2}$s$^{-0.5}$K$^{-1}$ have been estimated at 67P from Rosetta measurements, with variation between location on the nucleus, measuring instrument, local time of day, etc. \citep{Groussin2019}. The best constrained values typically come from areas at low temperature (e.g.~$10-60$ Jm$^{-2}$s$^{-0.5}$K$^{-1}$ from night side observations at $25-50$ K; \citealp{Choukroun2015}) so that, although our values with the fixed, high heat capacity are then somewhat high, especially for cm pebbles, they cannot be ruled out by the Rosetta data. In the following sections we test various values in our thermophysical model. For CO, we use a fixed heat capacity of $c_{CO}=884$ J kg$^{-1}$ K$^{-1}$, while its sublimation pressure and latent heat are as used in \cite{Gkotsinas2024}. We also note that large heat capacities and thermal inertias may help explain night-side activity which has been observed on 67P (e.g.~for roughly an hour by \citealp{Knollenberg2016}).

Gas diffusion is approximated by an analytical expression using the half-transmission thickness (or diffusion-scale parameter), $b$. From \citet{Gundlach.2020}, the local sublimation rate, $q,$ is given by
\begin{equation}
    q = P_{sat} \sqrt{\frac{m}{2 \pi k T}} \, \eta \alpha,
    \label{eq:qout}
\end{equation}
with the temperature-dependent saturation pressure, $P_{sat}$, and molecular mass,  $m$, of the relevant species. All physical parameters here are as in \citet{Attree2024b} and \citet{Gundlach.2020}. The outgassing area per layer, $\alpha$, is normalised to the equivalent flat surface area of the simulated region (one square metre), as found in laboratory experiments of porous ice outgassing, while 
\begin{equation}
    \eta = \left( 1 + \frac{z}{b} \right)^{-1}
    \label{eq:eta}
\end{equation}
describes the permeability. As in the previous works, half of the sublimating gas in each layer is assumed to flow outwards ($j_{leave}=q/2$), and is summed over all layers to find the total outgassing rate per time-step. The remaining gas is assumed to flow inwards ($j_{inward}=q/2$) and is then distributed as condensation over a number of deeper layers with the factors given in \citet{bischoff2023}. Also as before, we neglect re-condensation above the sublimation front as a minor contribution to the energy balance. The source and sink term $Q$ in Equation \ref{eq:heat_transport_eq} is then
\begin{equation}
    Q = \Lambda \, (j_{inward} - q) \, dt \,
    \label{eq:latent_heat}
\end{equation}
when summed over each species (H$_2$O, CO$_2$, and CO) with latent heat, $\Lambda$. Gas pressures for each volatile are computed using the damping term (eqn.~\ref{eq:eta}) and the sum compared to layer strength to determine if an ejection happens. Layer strength is considered as the sum of gravitational pressure and the tensile strength given by the depth/size dependent relation of \citet{Skorov.2012} (see discussion in section 2.3 of \citealp{Attree2024b} for a comparison of theoretical strengths with those measured at 67P: e.g.~\citealp{Biele2015, Spohn2015, Biele.2022b}). In this work we do not consider the internal pebble pressure model of \citet{Fulle2019}. 

The above model (equations \ref{eq:qout}, \ref{eq:eta}, and \ref{eq:latent_heat}) represents an approximation of the gas transport in-which the behaviour is determined by the $b$ parameter. This parameter, which is equal to the number of particle layers required to reduce the outgassing rate by a factor of two, is proportional to the particle size, $D_{p}$ \citep{Macher2024}. For gas-flow between the pebbles, this is the pebble size, $D_{p}=2R$. In section 2.4 of the previous paper \citep{Attree2024b} we described in detail how this proportionality varies with the material porosity, according to theoretical and direct simulation Monte Carlo work (DSMC). For effective porosities of $[0.4-0.8]$, $b\approx[0.3-3]D_{p}$, with a significant amount of scatter due to the uncertainty in the exact structure and arrangement of the particles (e.g.~hierarchical layering, surface scattering properties, polydispersity in particle sizes, etc.; see  \citealp{, Skorov.2021, Reshetnik.2022, Skorov.2022, Reshetnyk2025}). We therefore begin with $b=0.3D_{p}$, as the appropriate value for the 0.6 pebble packing fraction, but investigate varying this parameter in section \ref{sec:results.diffusivity} below. Preliminary tests with a diffusivity parameter of $b = 1D_{p}$ generated only sporadic ejections and no repeating activity, consistent with the results of \citet{bischoff2023} and \citet{Attree2024b}.

Insolation is calculated using the SPICE kernels for 19 different latitudes on a spherical comet with 67P's area-equivalent radius and orbit, and used in the standard energy balance formula with outgoing thermal radiation by the Stefan-Boltzmann law in the top layer. The bottom boundary is set to the constant, initial temperature (i.e.~$\frac{dT}{dt}=0$) which, motivated by the minimum surface temperatures seen in initial runs, we now set to 30 K rather than the 50 K used in the previous papers. We also performed test runs with 50 K and 75 K initial temperatures with little difference in results found. Three full orbital periods are simulated to ensure repeating activity, and the mass fluxes per latitude multiplied by their equal areas and summed to compare to the Rosetta observations. The other parameters are as described in the previous works.

\section{Results}
\label{sec:results}

For a diffusivity of $b = 0.3D_{p}$, repeated activity was detected over 3 orbital periods for many latitudes, with cycles of dry-crust formation interspersed with ejections by pressure build-up from sublimating H$_{2}$O or CO$_2$.

\begin{figure}
\resizebox{\hsize}{!}{\includegraphics[scale=0.5]{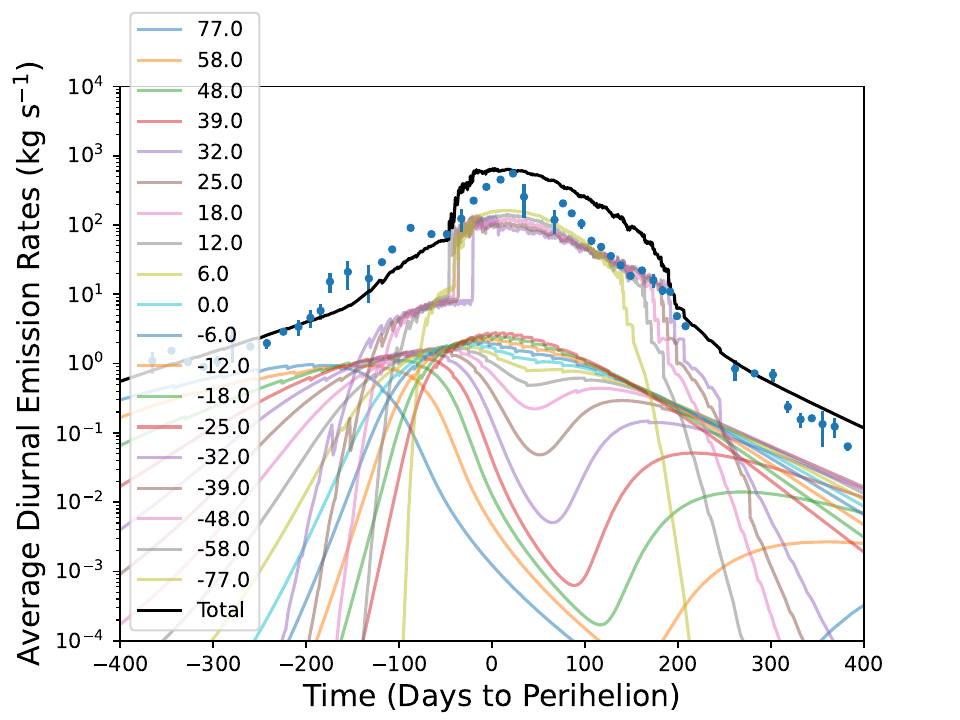}}%
\qquad
{\includegraphics[scale=0.5]{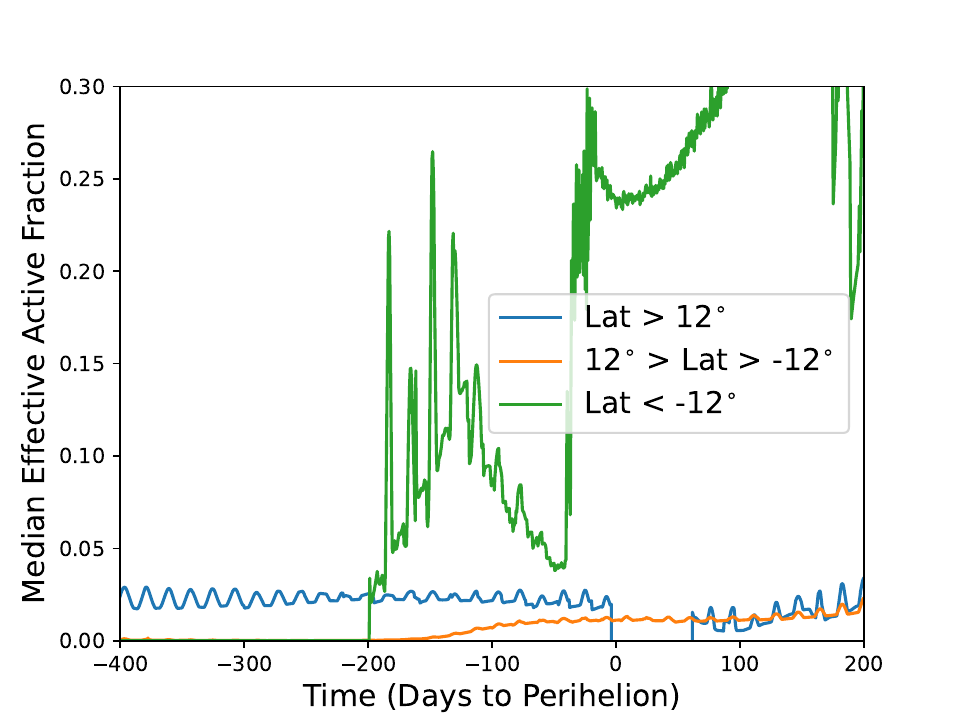}}%
\caption{Water production for a model with $\delta=2$, f$_{CO_{2}}=0.1$, f$_{CO}=0.01$, and $b=0.3D_{p}$. Top: mean diurnal water-production curve with time compared to the Rosetta data \citep{Laeuter2020} (solid points). Bottom: mean Effective Active Fraction (EAF), relative to a pure water-ice surface, across three latitude ranges.}
\label{plot_ResultsQ}
\end{figure}

Figure \ref{plot_ResultsQ}, top, shows the summed emission rates for a model run with our nominal parameter set of dust-to-total-ice-mass ratio $\delta = 2$, ice fractions $f_{CO_2}=0.1$ and $f_{CO}=0.01$, and pebble size $D_{p}=1$ cm. The water-production curve is reasonably well matched to the Rosetta data, albeit exceeding the measurements between $-50$ and $+200$ days of perihelion. The summed total water-emission over one orbital period is 1.6 times that reported by \citet{Laeuter2020}. Table \ref{Tab:results} lists the total water outgassed as well as the total ejected mass in other gas species and dust compared to the Rosetta observations (where the dust mass-loss is calculated from the total mass-loss value recently updated by \citealp{Laurent-Varin2024}). When examining the individual latitudinal curves, it can be seen that northern and equatorial regions dominate the water production up until around $-200$ days before perihelion, and again from around $275$ days after perihelion. Northern latitudes experience a dip in production around perihelion as they enter polar night. Equatorial regions maintain some outgassing, while latitudes south of $-18^{\circ}$ undergo a sharp rise in water-production as they become illuminated, followed by another sudden increase as dust ejections start, bringing the water closer to the surface. This 'activation' of southern latitudes by dust ejections leads to a much higher water-production rate than other latitudes. This can be visualised by calculating the effective active fraction (EAF) of the surface, relative to that of pure water-ice, as shown averaged over three latitude ranges in the bottom part of Fig.~\ref{plot_ResultsQ}. Mean EAF is always less than one as water-ice is only briefly present on the surface before retreating beneath a dust-crust where its emission is damped. The northern and equatorial latitude ranges maintain a relatively constant EAF of around a few percent (with a slight trend following the insolation), while the southern curve follows a more complicated trend. It first rises with increasing insolation as the southern hemisphere heats up, before falling slightly as water drains into the subsurface. Ejections then take over, bringing the water back close to the surface and leading to a high EAF. As insolation declines, gas emission does likewise, but much more slowly than in a simple surface energy-balance model due to the presence here of thermal inertia. This means that our calculated EAF continues to rise to a maximum of around 0.7. For the most relevant parts of the curves (when the respective latitudes are illuminated and outgassing is high), our EAF values of a few percent in the north and a steep rise to around $20-30\%$ in the south agree well with models fitted to 67P's non-gravitational accelerations (NGAs) and torques \citep{Attree2024, Attree2023, Attree2018}.

\begin{table*}
\caption{Input parameters and results for models with dust-to-total-ice-mass ratios, $\delta$, CO${2}$ and CO ice-mass fractions, f$_{CO_{2}}$ and f$_{CO}$. Results are normalised by the mass-loss over one orbit: observed $\Delta$M$_{H_{2}O}=4.5\times10^{9}$ kg, $\Delta$M$_{CO_{2}}=1.0\times10^{9}$ kg, $\Delta$M$_{CO}=1.9\times10^{8}$ kg \citep{Laueter2019}, and $\Delta M_{dust}=2.1\times10^{10}$ kg (\citealp{Laurent-Varin2024} minus the sum of all the volatiles in \citealp{Laueter2019}).}
\begin{tabular}{cccccccc}
\hline
Heat capacity, c (J kg$^{-1}$ K$^{-1}$) & $\delta$ & f$_{CO_{2}}$ & f$_{CO}$ & $\Delta$M$_{H_{2}O}$ & $\Delta$M$_{CO_{2}}$ &  $\Delta$M$_{CO}$ & $\Delta$M$_{dust}$ \\ \hline
Variable & 2 & 0.1 & 0.01 & 1.6 & 10.7 & 3.4 & 18.2 \\
Variable & 2 & 0.03 & 0.01 & 0.7 & 0.3 & 0.3 & 1.1 \\
Variable & 2 & 0.01 & 0.01 & 0.7 & 0.05 & 0.1 & 0.4 \\
Variable & 4 & 0.03 & 0.01 & 0.4 & 0.1 & 0.1 & 0.7 \\
Variable & 8 & 0.1 & 0.01 & 0.2 & 0.2 & 0.07 & 0.7 \\
Fixed, c$_{dust}=3000$ & 2 & 0.1 & 0.01 & 0.8 & 1.3 & 0.4 & 1.5 \\
Fixed, c$_{dust}=3000$, $b=0.1D_{p}$ & 2 & 0.1 & 0.01 & 2.1 & 5.1 & 1.5 & 7.7  \\
\end{tabular}
\label{Tab:results}
\end{table*}

\begin{figure*}
\subfloat{\includegraphics[scale=0.3]{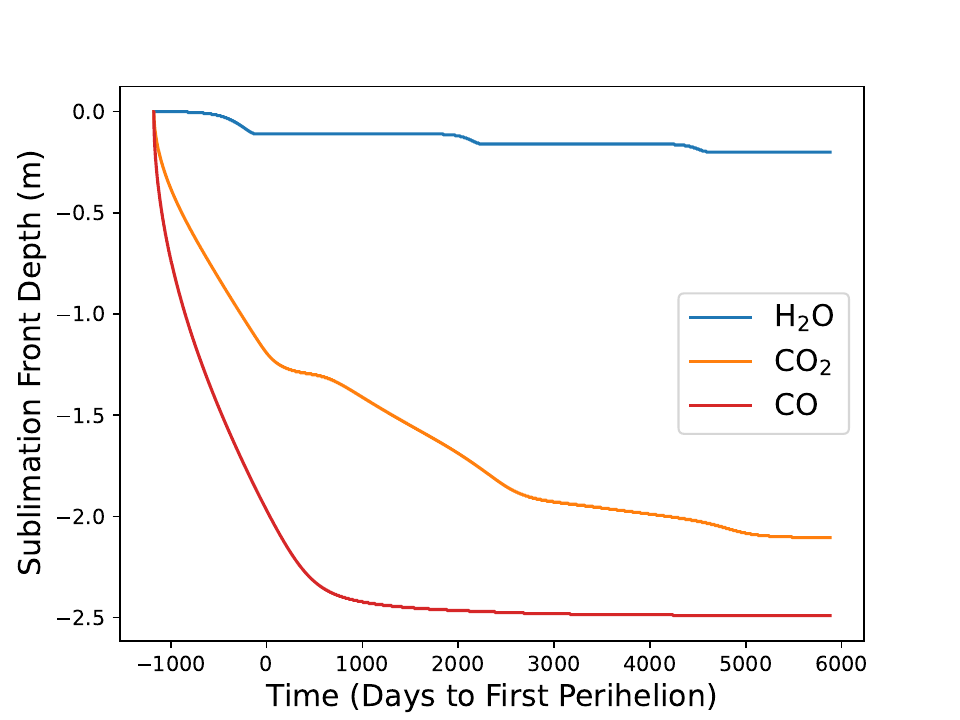}}%
\qquad
\subfloat{\includegraphics[scale=0.3]{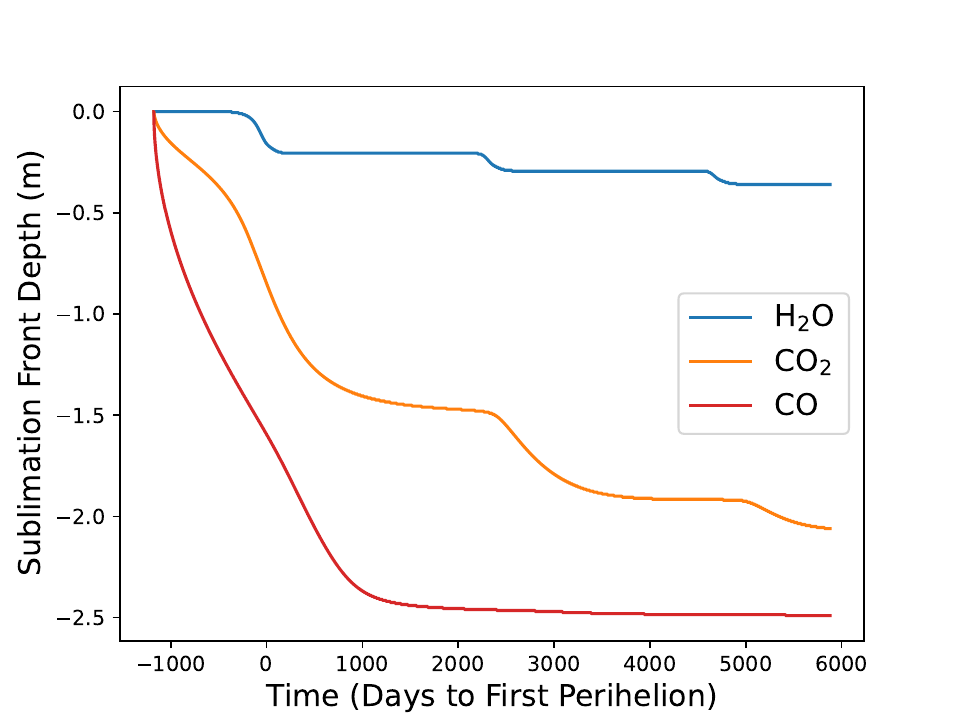}}%
\qquad
\subfloat{\includegraphics[scale=0.3]{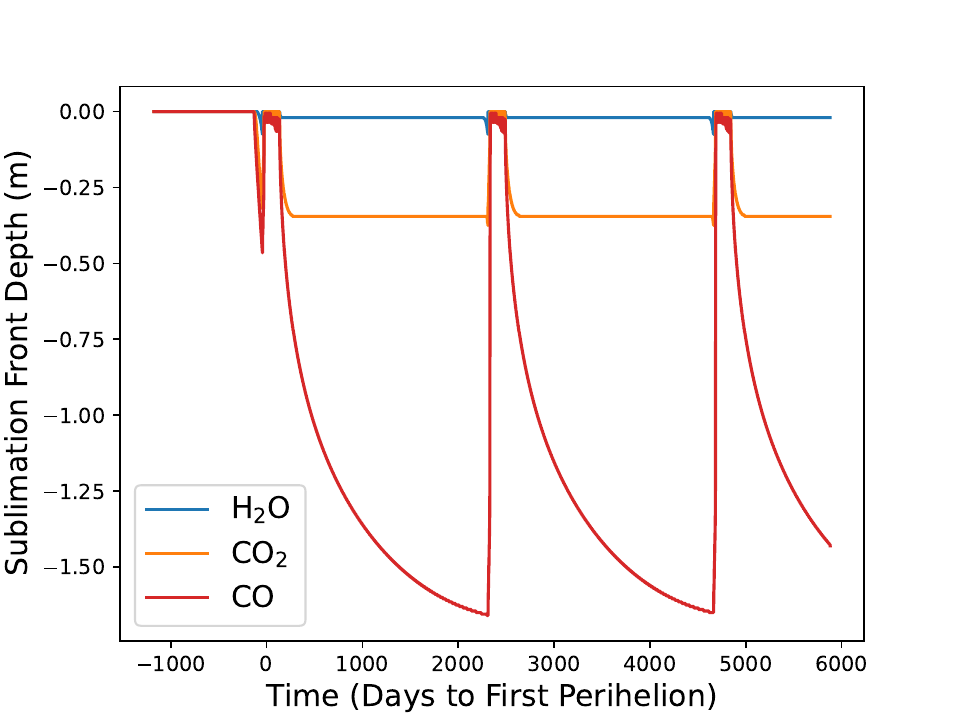}}%
\qquad
\subfloat{\includegraphics[scale=0.3]{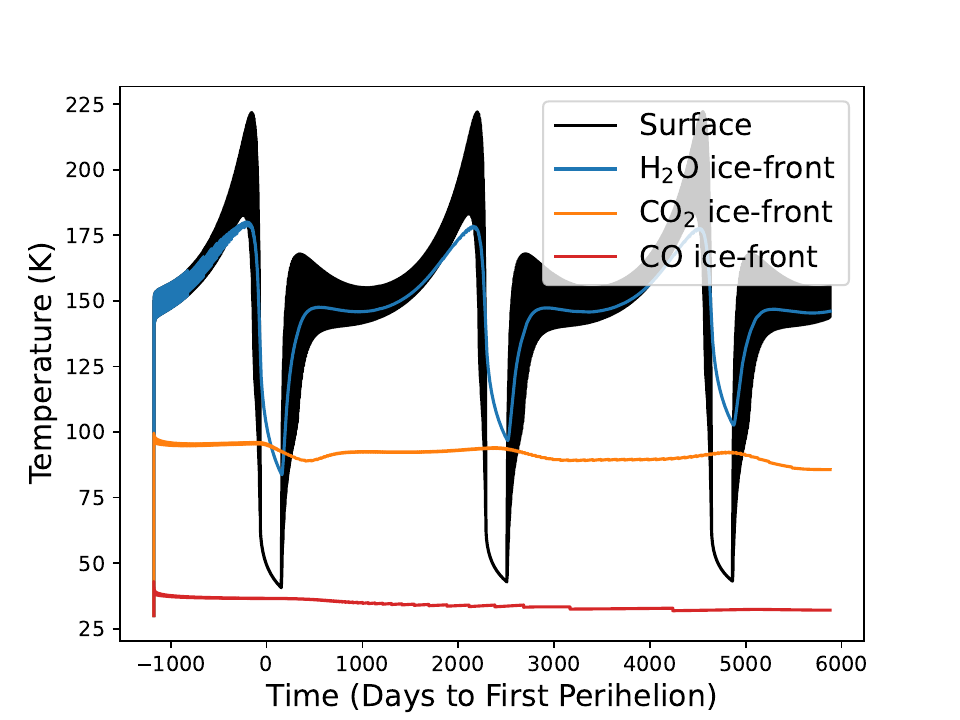}}%
\qquad
\subfloat{\includegraphics[scale=0.3]{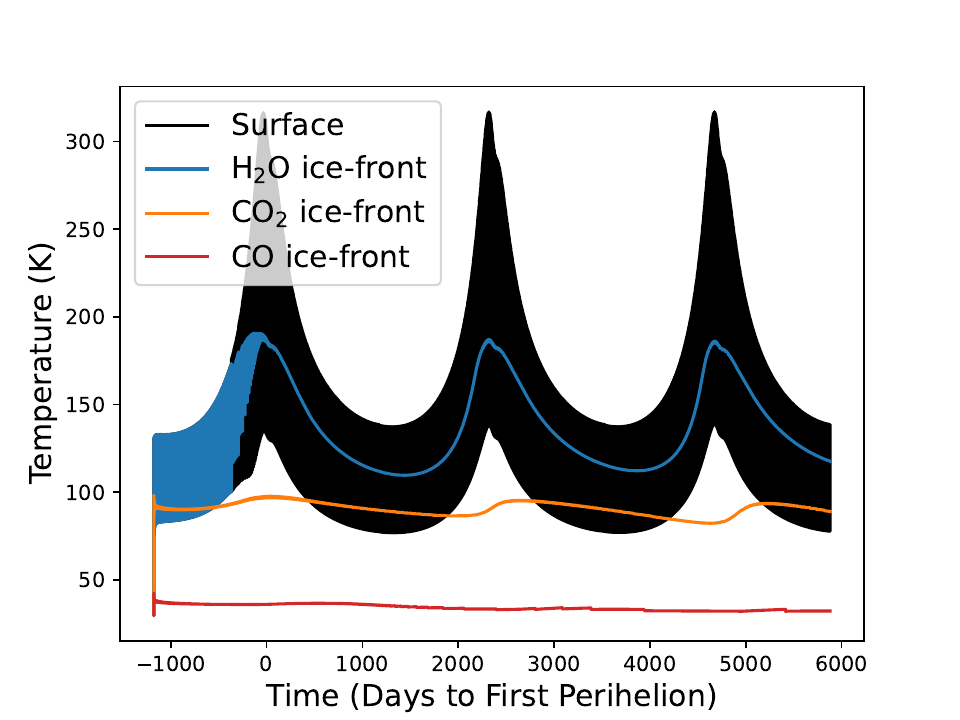}}%
\qquad
\subfloat{\includegraphics[scale=0.3]{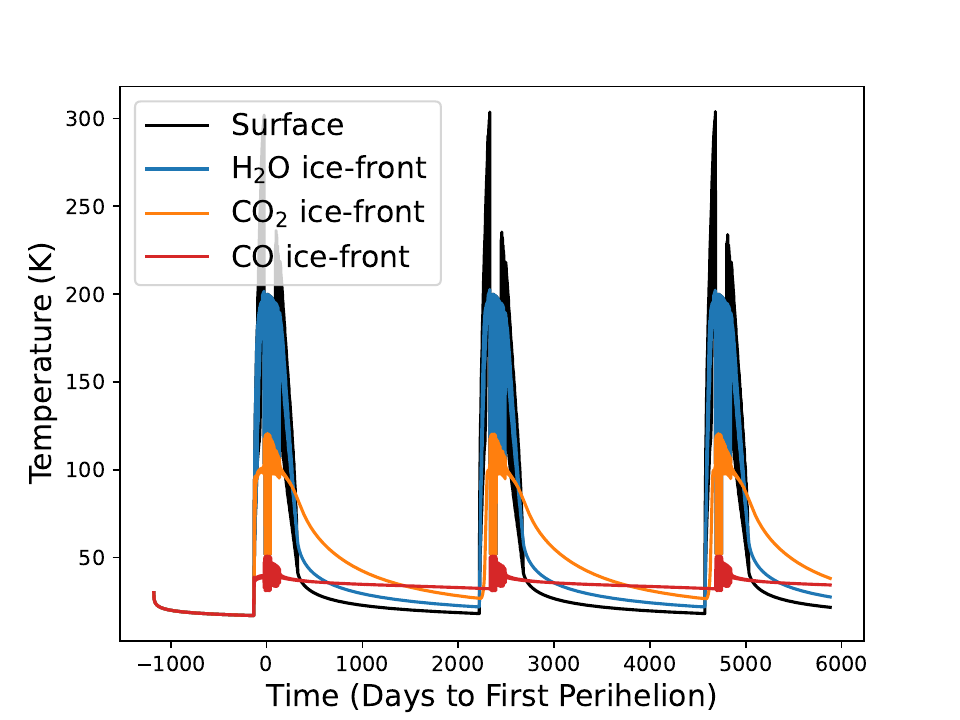}}%
\qquad
\caption{Sublimation front depths and temperatures for different latitudes in a model with $\delta=2$, f$_{CO_{2}}=0.1$, f$_{CO}=0.01$, and $b=0.3D_{p}$. Top: sublimation front depths for, from left to right, $+77^{\circ}$, $0^{\circ}$, and $-77^{\circ}$. Bottom: sublimation front temperatures.}
\label{plot_Results_sublimation fronts}
\end{figure*}

Figure \ref{plot_Results_sublimation fronts} shows the sublimation front depths (top) and temperatures (bottom) for the northern-most ($77^{\circ}$, left), equatorial (middle) and southern-most ($-77^{\circ}$, right) latitudes. Here we plot all three orbital periods to demonstrate the repeatability versus secular evolution of different latitudes. In the northern hemisphere and at the equator (left and middle), pressures never rise high enough to exceed tensile strength and the volatiles therefore continuously drain, reaching between tens of centimetres to several metres depth by the third orbit, for water and CO$_{2}$, respectively. CO drains further due to its low sublimation temperature, and in these runs exits the bottom of the 2.5 m deep simulation range and plays no further role in the northern hemisphere, consistent with its depth at several tens of metres in other thermal models (see e.g.~\citealp{Hoang2020}). The northern polar regions experience polar night around perihelion, leading to the pronounced dip in temperatures here. As pointed out by \citet{Groussin2024}, temperature inversions can occur between the $\sim100$ K interior and the surface, which rapidly cools to around 50 K before rising again after perihelion. At the equator, surface temperatures reach peak values at perihelion, ranging between $\sim130$ K and $>300$ K over the course of a day. The temperatures of the sublimation-fronts of the various volatiles are lower and with smoothed-out daily variations, corresponding to their large depths relative to the diurnal thermal skin-depth. 

Meanwhile, at high southern latitudes (right column) water is always near the surface, draining to a few centimetres depth during the outbound orbital leg and remaining there during southern polar night through aphelion. As insolation returns, close to the next perihelion, water drains to around 8 cm depth before triggering a chunk-ejection event. The surface is then active throughout perihelion, oscillating between water (and briefly even CO$_{2}$) being directly present on the surface, and draining below a thin dust-crust of only one or two pebble-layers thickness ($\sim0.5-2$ cm), which is then ejected again, predominantly by CO$_{2}$-pressure. Surface temperatures reach as low as $\sim30$ K at aphelion, with temperature inversions possible as the interior maintains some heat and temperatures of around $\sim40$ K. This is nonetheless much colder than the interior of the northern hemisphere ($\sim100$ K), a similar result to the thermal modelling of \citet{Groussin2024}. Peak surface temperature of around 300 K is reached during the brief period of water draining before ejections start. After this, frequent ejection events keep the surface from heating up much beyond the sublimation temperature of water-ice in the top pebble ($\sim200$ K), while the sublimation-front temperatures of the other volatiles oscillate up and down with their constantly changing depths.

\begin{figure}
\resizebox{\hsize}{!}{\includegraphics[scale=0.5]{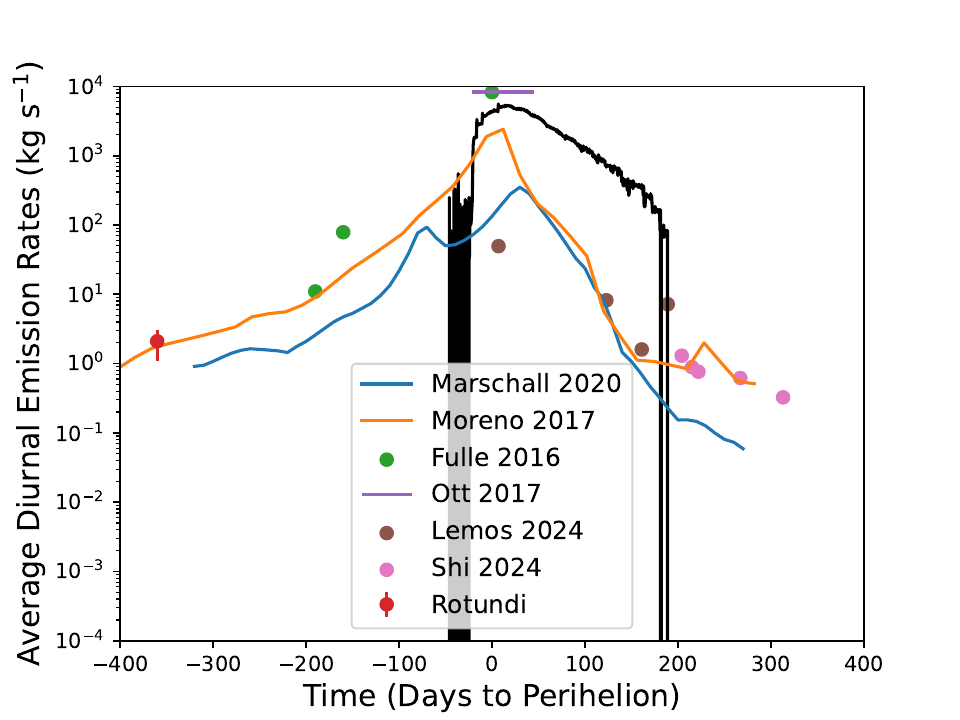}} %
\qquad
{\includegraphics[scale=0.5]{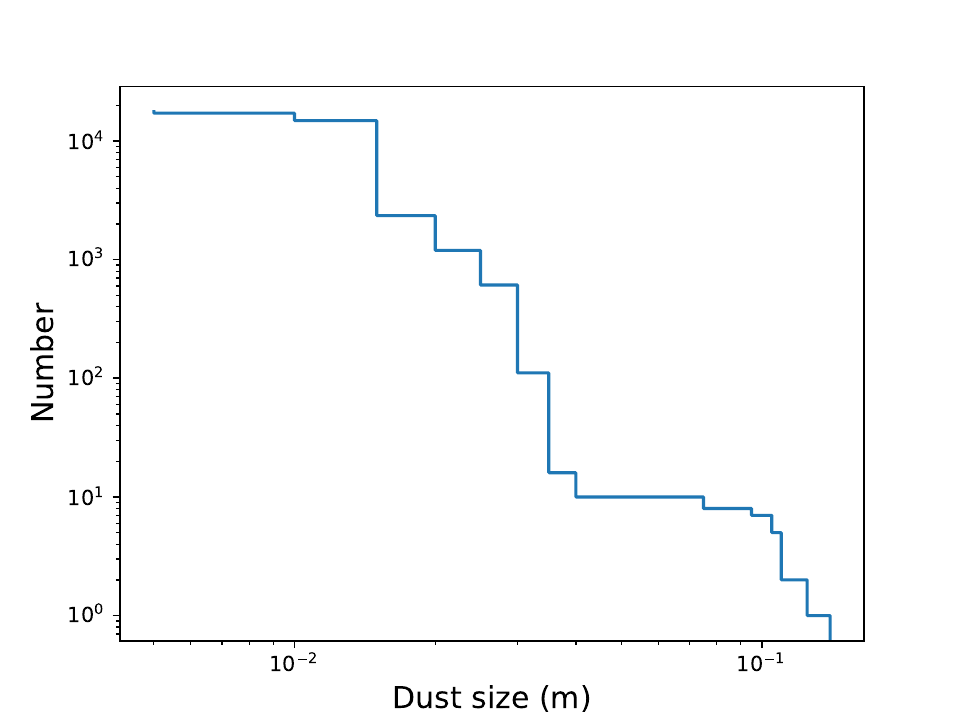}} %
\caption{Dust production for a model with $\delta=2$, f$_{CO_{2}}=0.1$, f$_{CO}=0.01$, and $b=0.3D_{p}$. Top: mean diurnal dust-production curve with time compared to the Rosetta data (coloured lines: \citealp{Marschall2020, Moreno2017}, solid points: \citealp{Fulle2016, Ott2017, Lemos2024, Shi2024, Rotundi2015}, truncated as described in the text). Bottom: size-frequency distribution of the modelled ejected dust. The exponent of a fitted power-law distribution is $-1.9$.}
\label{plot_ResultsQdust}
\end{figure}

Figure \ref{plot_ResultsQdust}, top, shows the time-dependent mass-flux of the ejected dust compared to a number of different observational estimates, while the bottom part shows the size-frequency distribution emitted over one whole orbit. The size of ejected particles is assumed to be equal to the depth where the ejection occurs, i.e.~their one-dimensional height. Ejected dust is then in the form of chunks or individual pebbles (ranging between $5$ mm and $\sim15$ cm-size here); sub-pebble dust is not emitted in this model. The ejections typically follow the above-noted pattern of one or more deep water-triggered ejections, followed by many small, CO$_{2}$-driven ones. Water-driven events eject dry dust particles, while those triggered by CO$_{2}$ can contain water-ice. We assume that the majority of this survives in fallback (e.g.~see \citealp{Davidsson2021b}) and so do not add this water to the total water-production rate. The size-frequency distribution of the ejected dust can be fitted by a power-law with an exponent of $-1.9$, which is shallower than that observed by Rosetta for small particles \citep{Moreno2017, Marschall2020} but not as shallow as some estimates for the larger chunks \citep{Ott2017}. When comparing the dust mass-fluxes in the top part of the plot, we only show the flux of particles larger than the minimum size ejected in the model. For the individual particle measurements \citep{Rotundi2015, Fulle2016, Ott2017} we plot only the relevant size bins, whereas for the continuous distributions \citep{Moreno2017, Marschall2020} we integrate the truncated power-laws. Our modelled dust-flux exceeds all the observations except those from tracking of individual particles by OSIRIS at perihelion \citep{Fulle2016, Ott2017}, while the total ejected mass is around 18 times too large compared to that estimated from the radio science experiment \citep{Laurent-Varin2024}. This may be because a large proportion of these, relatively big, chunks fall back to the surface and do not contribute to the lost mass. Alternatively, the whole southern surface may not be active. We do not model the irregular shape of 67P here, where steep cliffs and shadowed areas may reduce the energy input of certain areas below that of the threshold of activation by chunk ejections. We discuss these possibilities further in section \ref{sec:discussion} below. 

\begin{figure}
\resizebox{\hsize}{!}{\includegraphics[scale=0.5]{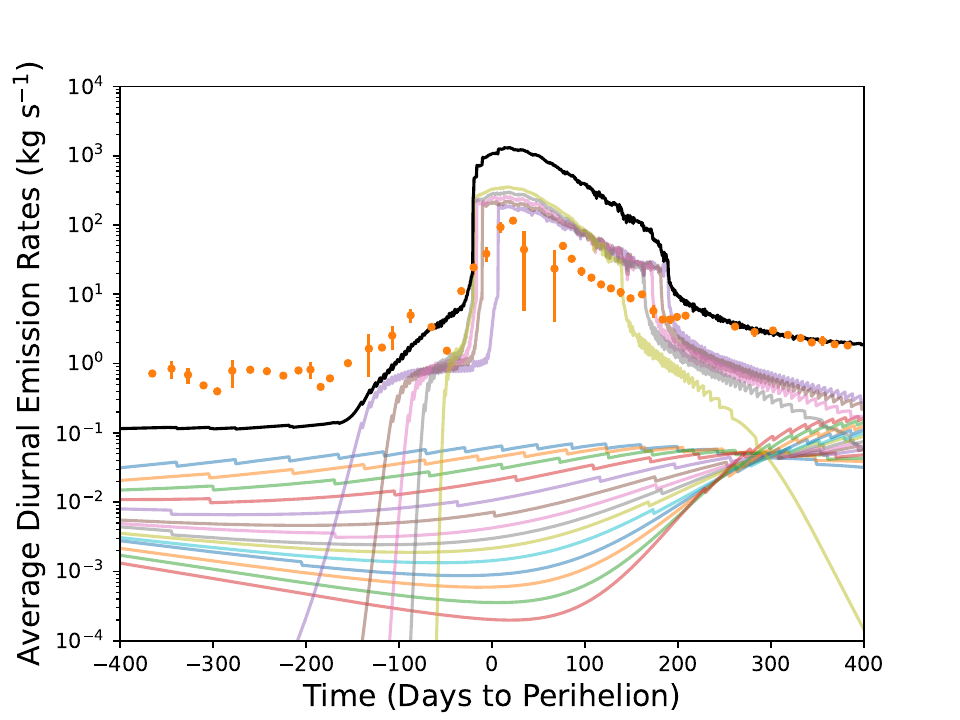}}%
\qquad
{\includegraphics[scale=0.5]{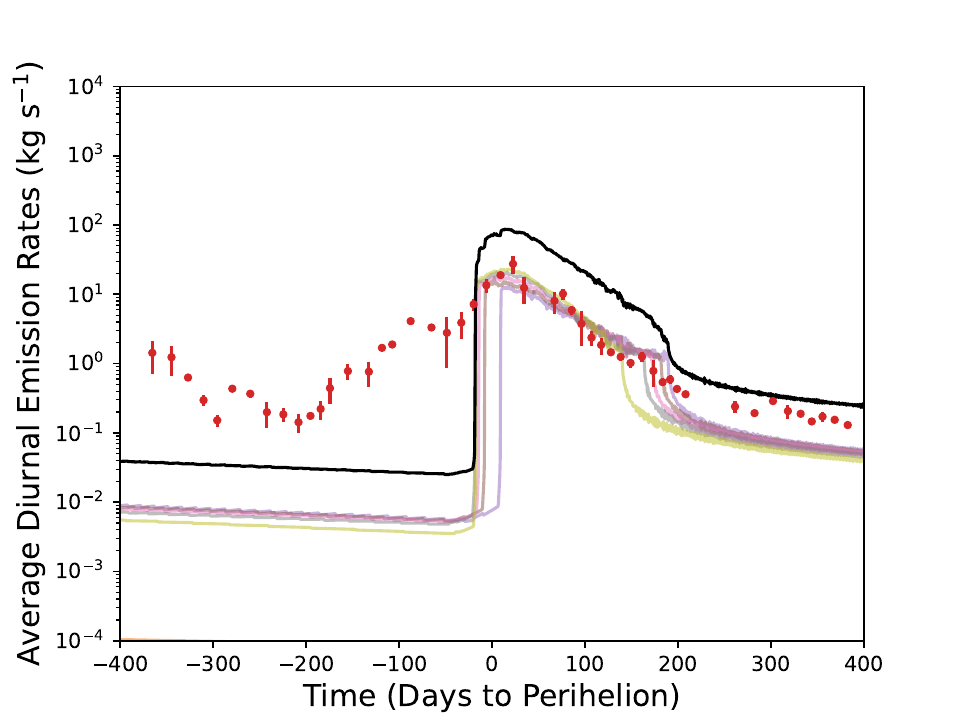}}%
\caption{Mean diurnal production rates for other gas species for a model with $\delta=2$, f$_{CO_{2}}=0.1$, f$_{CO}=0.01$, and $b=0.3D_{p}$, compared to the Rosetta data \citep{Laeuter2020} (solid points). Top: $CO_{2}$. Bottom: CO.}
\label{plot_ResultsQ_othervolatiles}
\end{figure}

In Figure \ref{plot_ResultsQ_othervolatiles}, we plot the emission rates of the two other modelled volatile species: CO and CO$_{2}$. In a similar way to dust emission, both rates are too high compared to the observations, especially around perihelion (approximately 3.4 and 10.7 times too much emitted total mass, respectively). Regarding the shapes of the curves, the post-perihelion period from $+200$ days seems well-matched in both cases (with the magnitude of CO being slightly too large, but with the correct shape). Here, ejections have ceased and the volatiles have swiftly drained into the subsurface (see Fig.~\ref{plot_Results_sublimation fronts}, top-right) and are cooling by radiation from the surface and conduction into the deep interior. The sublimation fronts for water and CO$_{2}$ have reach a constant depth (Fig.~\ref{plot_Results_sublimation fronts}, bottom-right) so that the sublimation rates are in balance with, and directly proportional to, the declining insolation, as also noted by \citet{Laeuter2020} for this period. Emissions of CO$_{2}$ and CO both predominantly come from the southern hemisphere at perihelion and post-perihelion, in agreement with the data \citep{Laueter2019, Hoang2019}. In contrast, before perihelion our modelled emission rates are too low, specifically before $-100$ days from perihelion, in the case of CO$_{2}$, and before perihelion itself, in the case of CO. For CO, this can be explained by the draining described above, which means that the whole northern hemisphere is now no-longer outgassing this species. We therefore miss the observed northern emissions as well as the decrease in CO emission with decreasing heliocentric distance seen before $-200$ days. CO$_{2}$, by contrast, comes primarily from the northern hemisphere during this period, which does not agree with numerous Rosetta observations \citep{Laueter2019, Hoang2019} showing the bulk of its emission from the south during all Rosetta mission phases. Thus, we are also missing some southern hemisphere CO$_{2}$ emission. The results of this model therefore, although promising, show some discrepancies with the data that we will now investigate by varying a number of input parameters.

\subsection{Ice content}

Since our emitted CO and CO$_{2}$ rates do not fit the data, we first investigate how varying the total and relative ice fractions change the results. Table \ref{Tab:results} lists the total emitted masses for various parameter combinations. Lowering the overall ice fraction (increasing $\delta$, which means reducing both water-, CO$_{2}$-, and CO- mass per layer) leads to a reduction in outgassing of all species, as expected. The smaller mass of ice in each depth-step leads to it draining out faster, and therefore to deeper sublimation depths and reduced fluxes. Deeper sublimation fronts struggle to generate enough pressure to eject layers, leading also to a reduction in the dust mass-flux. Experiments with $\delta=50$ \citep{Ciarniello2022} drained completely, showing no dust-ejection activity.

For the results with a fixed total ice-mass-ratio but decreasing CO$_{2}$ fraction, a similar trend is seen.  Naturally, this is particularly pronounced for CO$_{2}$ itself, but this species is also very important in driving ejections during the active phase when it is close to the surface. Less CO$_{2}$ therefore means fewer ejections, leading to a reduced dust-mass-flux, and also to reduced water outgassing as the dust-crust is on average thicker than before. For very low CO$_{2}$ contents, only water-driven ejections remain, with an ejection rate that is too low to match the data. Thus, both water-ice and CO$_{2}$-ice contents are critical in determining the activity behaviour. When varying the CO$_{2}$ fraction, we do not find a balance where the flux matches the observed outgassing rates: either CO$_{2}$ drives very frequent ejections, and thus remains near the surface leading to very high emission rates, or it drains away, leading to very low emission rates and low dust fluxes. This is a similar problem to that faced in \citet{Attree2024b} when simulating the non-WEB material.

\subsection{Heat capacity}

\begin{figure}
\resizebox{\hsize}{!}{\includegraphics[scale=0.5]{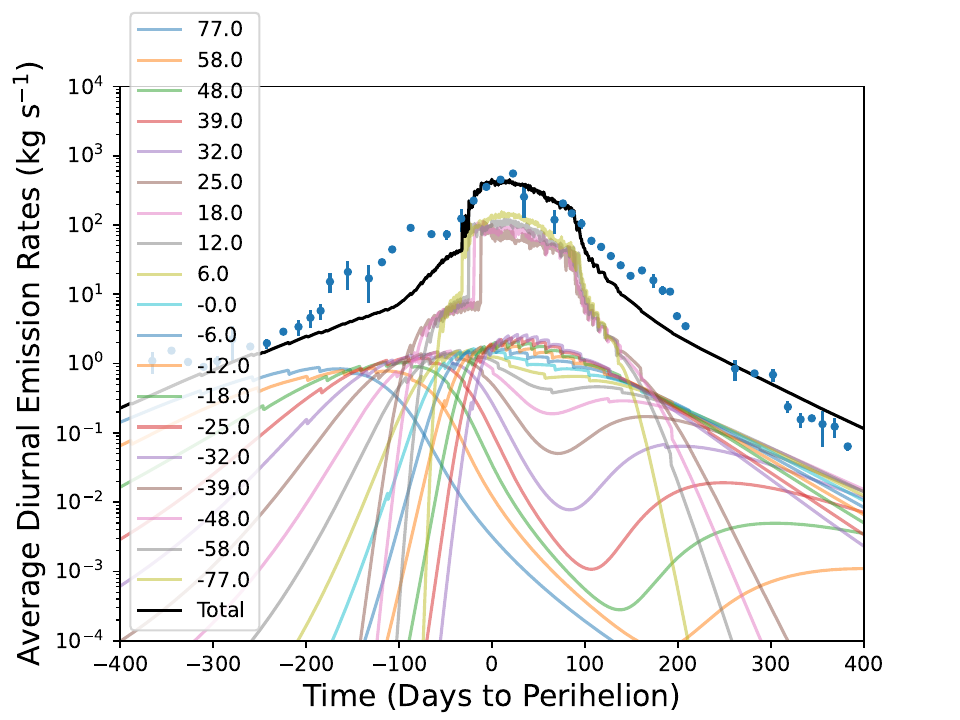}}%
\qquad
{\includegraphics[scale=0.5]{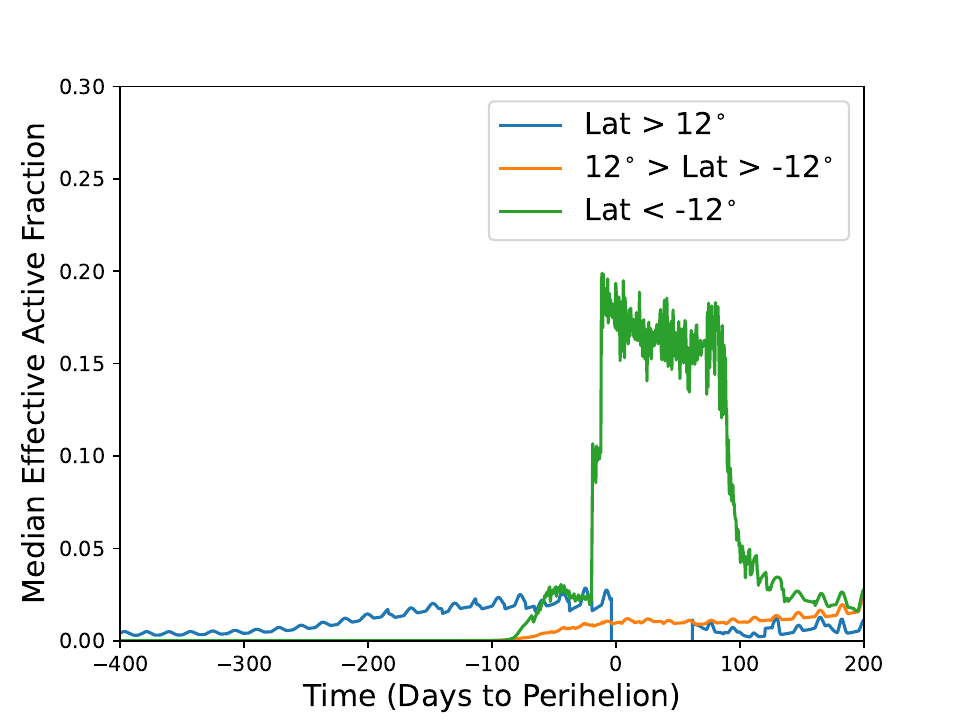}}%
\caption{Water production for a model with $\delta=2$, f$_{CO_{2}}=0.1$, f$_{CO}=0.01$, $b=0.3D_{p}$, and a fixed dust heat-capacity. Top: mean diurnal water-production curve with time compared to the Rosetta data (solid points). Bottom: mean Effective Active Fraction (EAF), relative to a pure water-ice surface, across three latitude ranges.}
\label{plot_ResultsQ_c3000}
\end{figure}

\begin{figure}
\resizebox{\hsize}{!}{\includegraphics[scale=0.5]{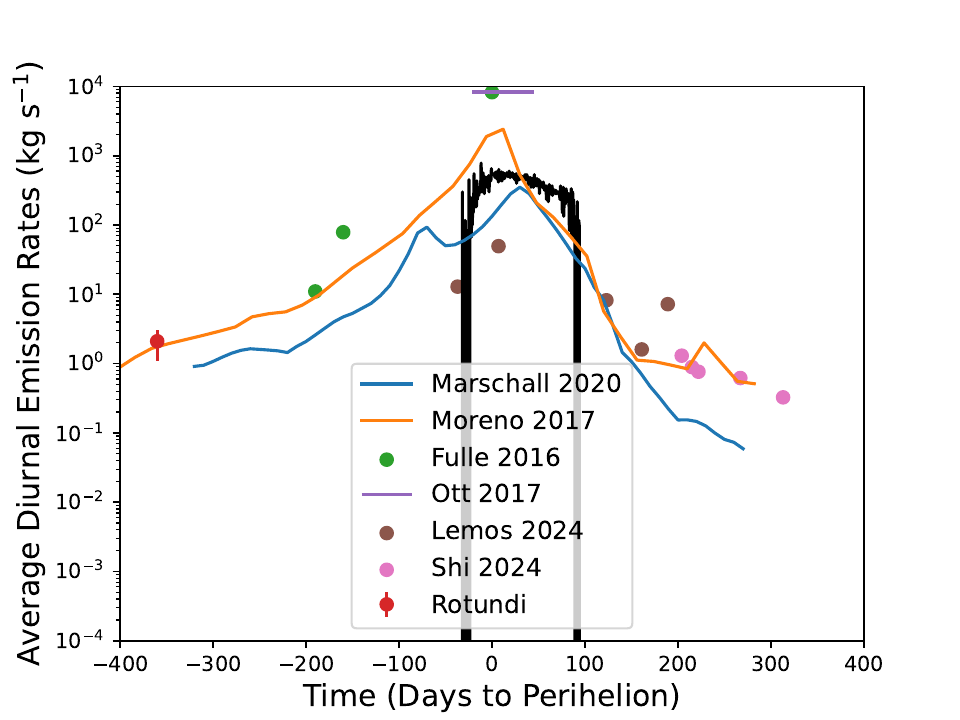}} %
\qquad
{\includegraphics[scale=0.5]{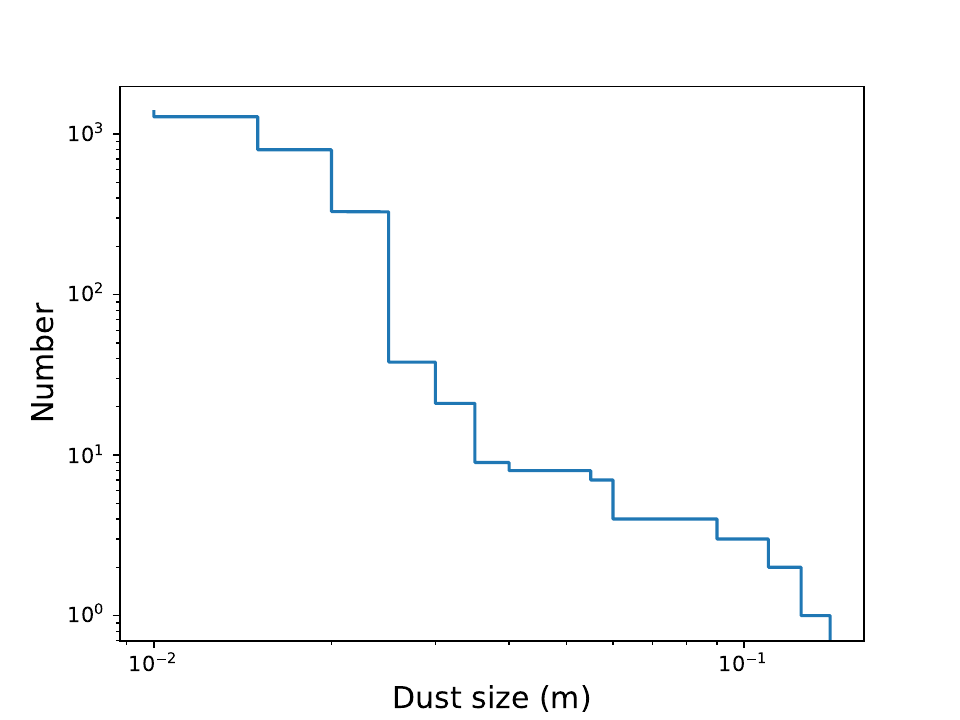}} %
\caption{Dust production for a model with $\delta=2$, f$_{CO_{2}}=0.1$, f$_{CO}=0.01$, $b=0.3D_{p}$, and a fixed dust heat-capacity. Top: mean diurnal dust-production curve with time compared to the Rosetta data (coloured lines and solid points). Bottom: size-frequency distribution of the modelled ejected dust. The exponent of a fitted power-law distribution is $-2.6$.}
\label{plot_ResultsQdust_c3000}
\end{figure}

\begin{figure}
\resizebox{\hsize}{!}{\includegraphics[scale=0.5]{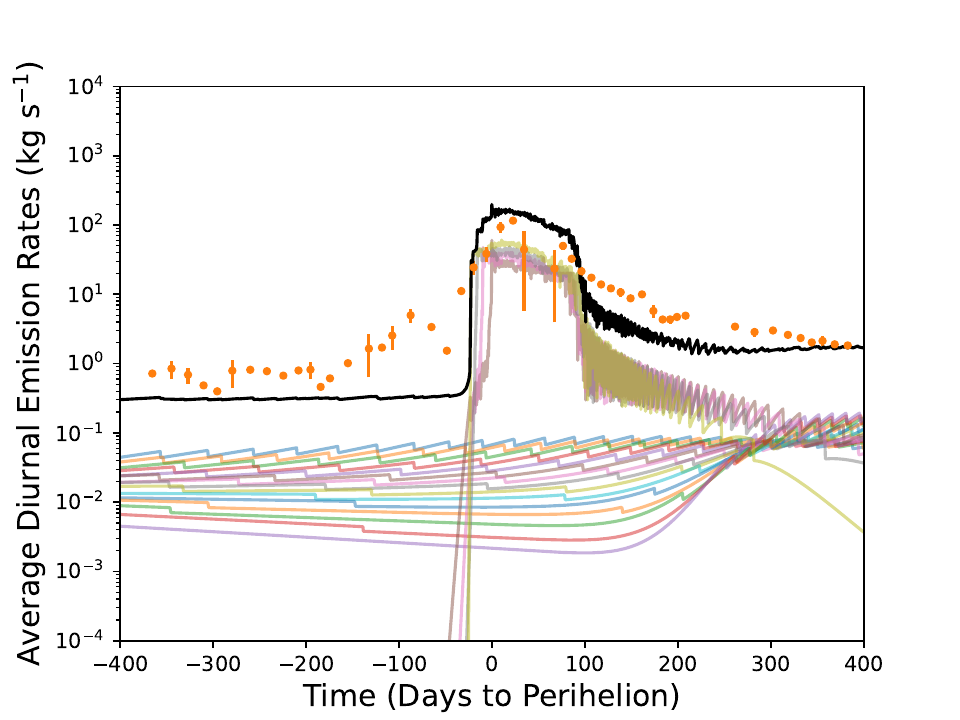}}%
\qquad
{\includegraphics[scale=0.5]{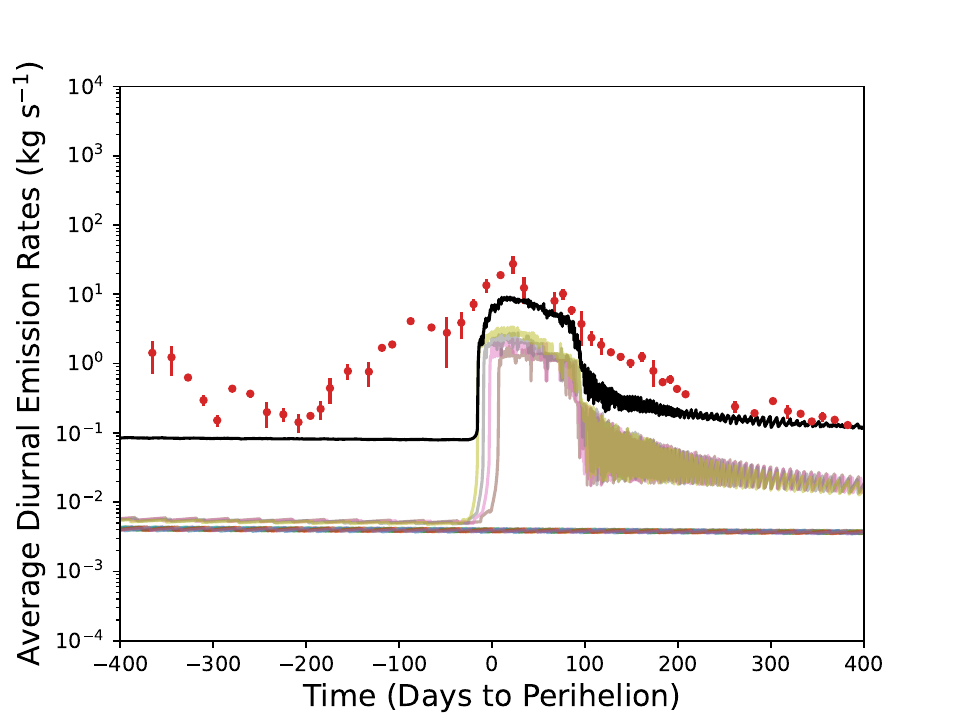}}%
\caption{Mean diurnal production rates for other gas species for a model with $\delta=2$, f$_{CO_{2}}=0.1$, f$_{CO}=0.01$, $b=0.3D_{p}$, and a fixed dust heat-capacity, compared to the Rosetta data (solid points). Top: $CO_{2}$. Bottom: CO.}
\label{plot_ResultsQ_othervolatiles_c3000}
\end{figure}

We next tested the model with the fixed, and high, heat-capacity of dust, $c_{dust}=3000$ J kg$^{-1}$ K$^{-1}$, used in \citet{Attree2024b}. As can be seen by the results in table \ref{Tab:results} and figures \ref{plot_ResultsQ_c3000}, \ref{plot_ResultsQdust_c3000}, and \ref{plot_ResultsQ_othervolatiles_c3000}, the mass fluxes of the various species and dust are now a closer match to the observations. This is because the larger heat-capacity of the material means that each layer takes longer to heat up, delaying the sublimation and draining of ices in the inactive, northern latitudes, and reducing the frequency of CO$_{2}$-driven ejection events around perihelion for the active latitudes. Thus, the super-volatiles, CO and CO$_{2}$, have higher fluxes away from perihelion and lower fluxes around perihelion, more closely matching the Rosetta data. The water flux, meanwhile, is slightly lower than the observations due to the decreased erosion rate leaving it deeper than before on average. The changed balance between erosion and draining also means that fewer latitudes reach enough pressure to generate ejections so that activity is limited to latitudes of $-39^\circ$ and below in this model. The CO and CO$_{2}$ production curves leading up to perihelion and from 100 days after are then worse fits to the data than in the previous model, due to the more sharp transition from no activity to activity found with only a few active southern latitudes.

\subsection{Diffusivity}
\label{sec:results.diffusivity}
\begin{figure}
\resizebox{\hsize}{!}{\includegraphics[scale=0.5]{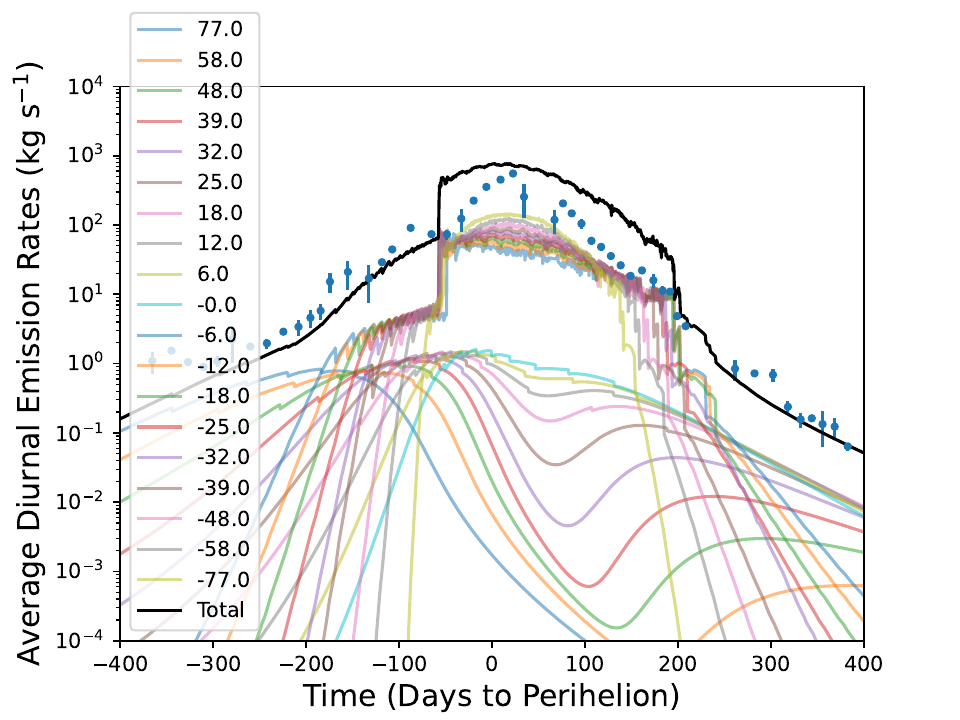}}%
\qquad
{\includegraphics[scale=0.5]{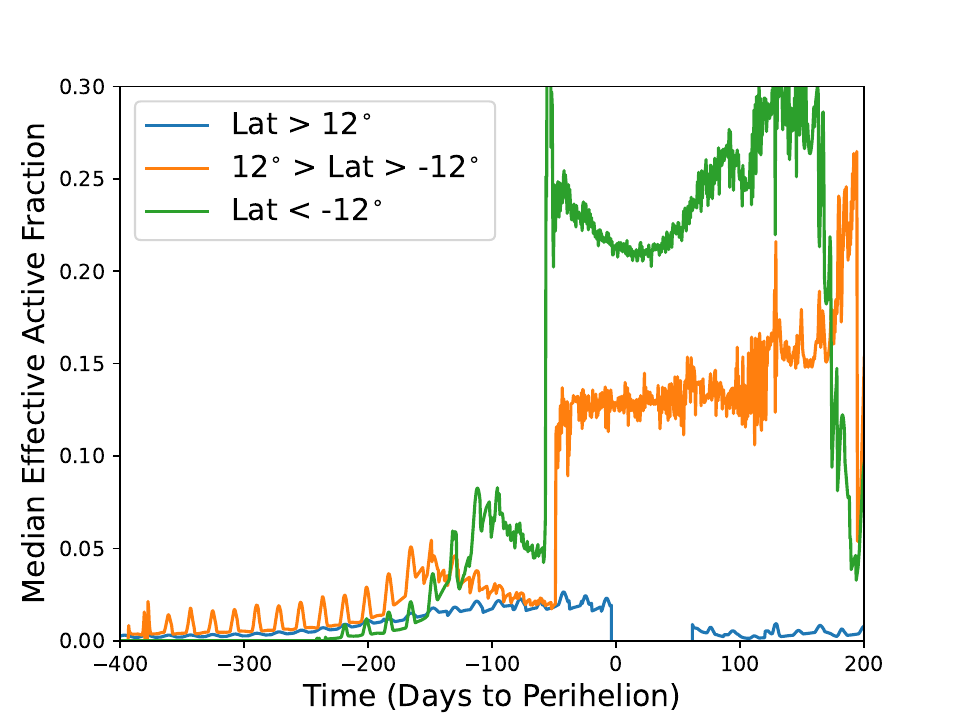}}%
\caption{Water production for a model with $\delta=2$, f$_{CO_{2}}=0.1$, f$_{CO}=0.01$, $b=0.1D_{p}$, and a fixed dust heat-capacity. Top: mean diurnal water-production curve with time compared to the Rosetta data (solid points). Bottom: mean Effective Active Fraction (EAF), relative to a pure water-ice surface, across three latitude ranges.}
\label{plot_ResultsQ_cvariable_b0.1}
\end{figure}

\begin{figure}
\resizebox{\hsize}{!}{\includegraphics[scale=0.5]{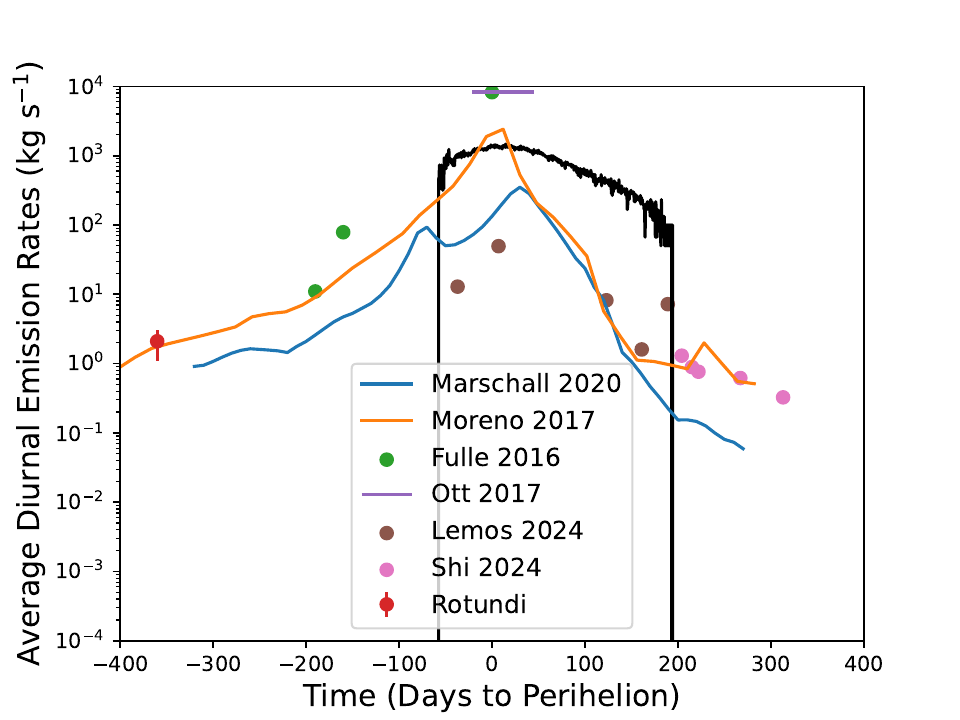}} %
\qquad
{\includegraphics[scale=0.5]{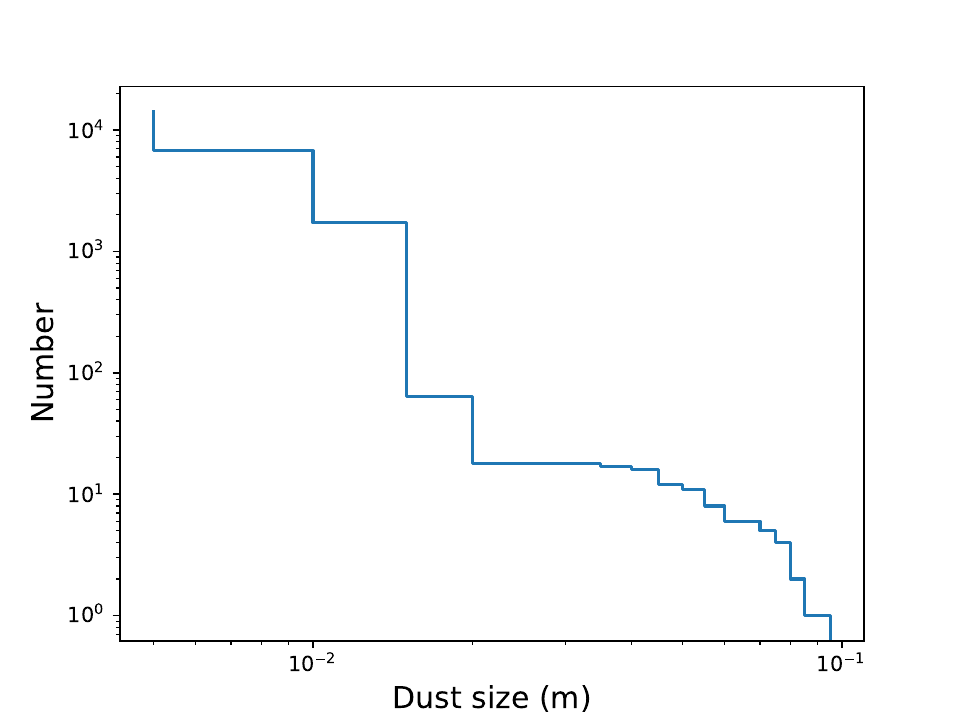}} %
\caption{Dust production for a model with $\delta=2$, f$_{CO_{2}}=0.1$, f$_{CO}=0.01$, $b=0.1D_{p}$, and a fixed dust heat-capacity. Top: mean diurnal dust-production curve with time compared to the Rosetta data (coloured lines and solid points). Bottom: size-frequency distribution of the modelled ejected dust. The exponent of a fitted power-law distribution is $-3.6$.}
\label{plot_ResultsQdust_cvariable_b0.1}
\end{figure}

\begin{figure}
\resizebox{\hsize}{!}{\includegraphics[scale=0.5]{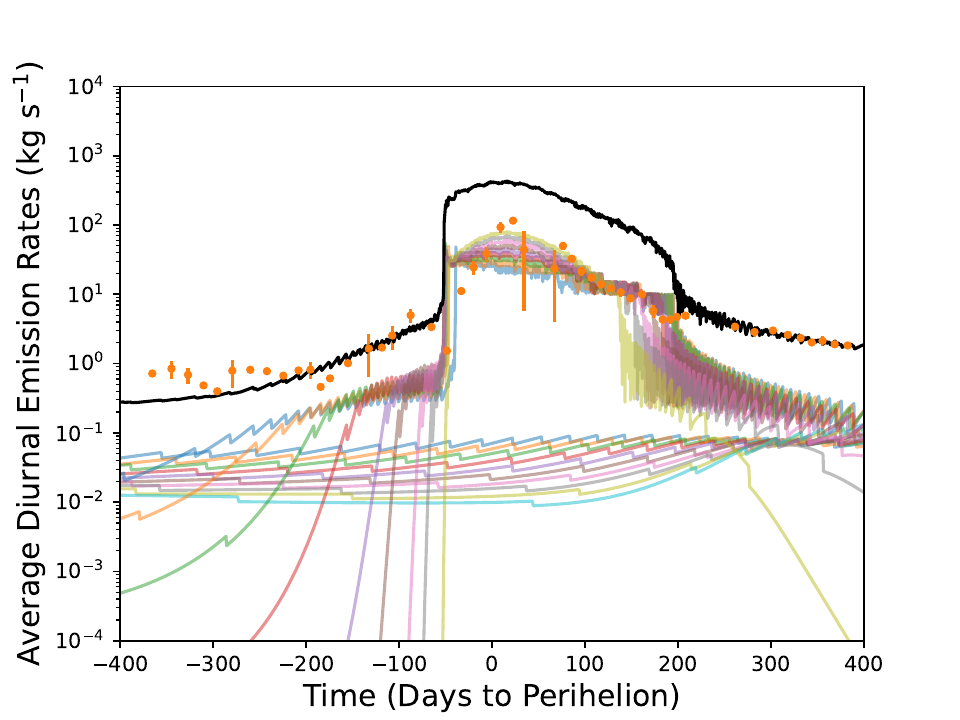}}%
\qquad
{\includegraphics[scale=0.5]{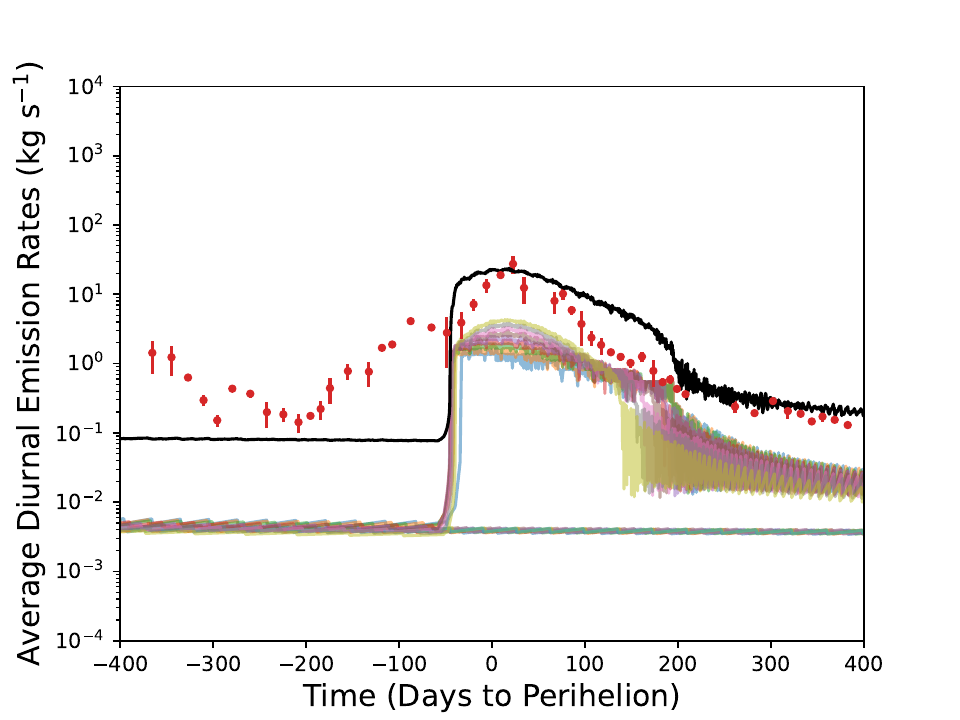}}%
\caption{Mean diurnal production rates for other gas species for a model with $\delta=2$, f$_{CO_{2}}=0.1$, f$_{CO}=0.01$, $b=0.1D_{p}$, and a fixed dust heat-capacity, compared to the Rosetta data (solid points). Top: $CO_{2}$. Bottom: CO.}
\label{plot_ResultsQ_othervolatiles_cvariable_b0.1}
\end{figure}

As discussed above, the diffusivity  parameter, $b$, should be determined by the dust structure and porosity but, since there are significant unknowns in both of these quantities at the local scale, we now examine how varying it can affect the simulation results. Larger $b$ values allow gas to more easily escape through the diffusive dust-crust, reducing pressure and ejections. For large enough values ($b\gtrsim0.6D_{p}$), only very sporadic ejections are observed. We test a smaller value of $b=0.1D_{p}$ to see the effect of reduced diffusivity. For the variable heat-capacity case, this results in very large dust and CO$_{2}$ fluxes as the reduced diffusivity leads to large pressure build-up and frequent ejections. Frequent ejections keep CO$_{2}$ close to the surface, resulting in very low temperatures and a correspondingly low heat-capacity, exacerbating the ejection-frequency problem. For a case with a low diffusivity as well as a high, fixed heat-capacity, the results, as shown in table \ref{Tab:results} and figures \ref{plot_ResultsQ_cvariable_b0.1}, \ref{plot_ResultsQdust_cvariable_b0.1}, and \ref{plot_ResultsQ_othervolatiles_cvariable_b0.1}, show a reduced flux in all the gas species compared to the previously stated model. This is because the larger heat-capacity slows the heating of near surface layers and damps the rate of these ejections. The total ejected masses are now closer to the Rosetta observations, although the mass fluxes of all gas species and dust are still rather large at perihelion. Whilst the higher heat-capacity helps with the peak ejection rates, the reduced diffusivity means that the sublimation fronts are closer to the surface on average than for higher diffusivity cases. The whole southern hemisphere can now reach sufficient pressures to overcome tensile strengths and trigger ejections. Better agreement with the shapes of the CO and CO$_{2}$ curves is then reached due to this more smooth distribution of activity. Additionally, CO is now present in the north at depths within 2.5 m of the surface, meaning it contributes some flux before perihelion, although the decrease prior to $-200$ days is still not seen. CO$_{2}$ is also closer to the surface, especially in the active south, meaning it contributes to an enhanced flux before perihelion. The crossover point from northern hemisphere to southern hemisphere dominated CO$_{2}$ emission now happens at around $-200$ days, while some low latitudes in the south actually produce the most flux all the way out to Rosetta's arrival date, in better agreement with the data. Finally, figure \ref{plot_ResultsQdust_cvariable_b0.1} (lower part) shows that the shallower sublimation fronts result in smaller ejected particles on average, with maximum size now $<10$ cm.

\subsection{Pebble size and numerical resolution}
\label{sec:results:resolution}

As previously mentioned, we tested our nominal model with a one-second time-step and found no significant differences. We also wished to investigate how changing the depth resolution effected the results. Here, the situation is somewhat complicated by the fact that normally our depth-step is tied to the assumed pebble radius so that changing the numerical resolution of the model also has the effect of changing pebble size, with all implications for material properties that that entails. To disentangle these effects, we first relaxed this association and tested a single-orbit run with centimetre pebbles resolved by a depth-step of 1 mm (and a time-step of $dt=1$ s to comply with \citealp{Patankar}). This means we are averaging material properties across a finite volume smaller than the structural size of the component particles, which may not result in an accurate physical description of the medium. Nonetheless, we proceed here in order to test the limitations of the model.

In a run of our nominal model ($\delta=2$, f$_{CO_{2}}=0.1$, f$_{CO}=0.01$, $b=0.3D_{p}$), we found qualitatively the same results as before. Ejections and large outgassing rates of all ice species are found in the southern hemisphere, while draining and a lack of dust-activity are noted at the equator and in the north. The minimum ejected particle-size decreased from 5 mm to 3 mm (resulting in a different size-distribution), and was driven by CO$_{2}$, with water still only ejecting 5 mm or larger particles, due to the high strength at smaller scales. The overall erosion rate increased significantly, due to the high frequency of CO$_{2}$-driven ejections close to the surface, meaning the results are even further from the data.

Finally, we tested a model with $dz=R=1$ mm, i.e.~two-millimetre diameter pebbles, and a time-step of one second. Again, we found very similar results with qualitatively the same behaviour. The increased strength close to the surface is compensated by the reduced diffusivity raising the pressure to the point where frequent CO$_{2}$-driven ejections lead to a very high erosion rate. The smallest ejected particles were again 3 mm in size, and no ejections were found at the equator or northern hemisphere.

Following this, we can say that changing both pebble size and spatial resolution does not effect the overall behaviour of our model. Nonetheless there are significant complexities at small scales near the surface and/or the sublimation front that can make a difference to specific results. Further work resolving the sublimation patterns within individual pebbles at the surface (which may require 3D modelling) is needed, but is beyond the scope of this paper.

\section{Discussion}
\label{sec:discussion}

\subsection{Outgassing patterns}
\label{sec:discussion:outgassing}

Our model of an ejecting dust-crust can roughly reproduce the global production rates of water, CO$_{2}$, CO, and dust particles larger than the pebble-size for a reasonable parameter set. The water outgassing distribution, which peaks strongly in the south with a relative activity level of EAF $\sim20\%$ compared to only a few percent in the north, provides a natural explanation for the results of non-gravitational acceleration and torque modelling \citep{Attree2024}. The results above favour low global dust-to-ice ratios ($\delta\sim2$), low CO and CO$_{2}$ fractions, as well as low diffusivities and/or large heat-capacities for best agreement with the Rosetta data. 

Low diffusivities are a natural consequence in the comet-nucleus model proposed by \citet{Fulle2017}. In this model, the nucleus of a comet is composed of densely packed pebbles, with the interstitial spaces filled with fractal dust particles. This model effectively accounts for the simultaneous release of both densely packed and highly porous dust aggregates, as measured in situ by the Rosetta mission's dust instrument, GIADA \citep{Fulle2015}. The comet nucleus features centimetre-sized fluffy particles with approximately 99\% porosity, consisting of micrometre-sized monomer grains. These particles significantly reduce the mean free path of gas molecules by about two orders of magnitude, down from centimetres (typical of void space devoid of particles) to approximately 100 micrometres when fluffy dust is interspersed among the pebbles. It is important to note that, although the diffusivity is reduced, the presence of extremely-porous dust does not have a comparable effect on thermal radiation, particularly when the wavelengths are much larger than the grain size. Thus, the heat conductivity remains largely unaffected despite the presence of porous dust.

We also note that our preference for low diffusivities, as well as low dust-to-ice ratios, agrees with the modelling of \citet{Davidsson2022}, using the NIMBUS software \citep{Davidsson2021MNRAS}. Our model produces different sublimation-front depths to that of \citet{Davidsson2022}, with water deeper in the north and CO$_{2}$ generally shallower, whilst our best-matching $10\%$ CO$_{2}$-fraction is somewhat lower that theirs. Some of this can be attributed to the differing diffusivities in the two models (void spaces here are assumed proportional to our large 1 cm pebbles, but are generally tens or hundreds of microns in \citealp{Davidsson2022}), but the majority of the disagreement stems from the different modelling of erosion. \citet{Davidsson2022} suppose that CO$_{2}$-driven chunk-ejection is a localised phenomenon and that the majority of the surface is progressively being eroded into small dust with an \textit{a priori} prescribed rate proportional to water emission. We have not so-far considered the erosion of fine dust, which will be discussed below in section \ref{sec:discussion:mechanism}, but note that this low level of erosion over the whole comet keeps water close to the surface everywhere. Meanwhile, our explicit modelling of pebble and chunk ejection is sensitive to CO$_{2}$ pressure, necessitating its relatively low abundance (in order to minimise its high ejection efficiency), and resulting in it being closer to the surface on average. Thus, the presence and location of erosion (into small dust or large chunks, and continuously everywhere or highly localised) is critical in determining the thermal balance and derived thermophysical properties of the cometary material.

Turning now to specific parts of the emission curves, we note that at perihelion we have very high mass-fluxes of CO, CO$_{2}$ and dust. This may imply that the whole southern surface is not active. However, table \ref{Tab:results} shows that the total mass-loss figures (which are dominated by emission at perihelion) are in excess of the measurements by different amounts for different species. It is hard to see how to scale down the dust, CO, and CO$_{2}$ emitting areas, for example, without also producing a water flux that is too low. For the dust, it could be argued that a large fraction of these pebbles and chunks should fall back to the surface, and thus not count towards the total mass-loss. This would improve the situation but would not help with the other species, which would still be too high by various factors. In general, it seems that dust-lifting activity, at least around perihelion, is something of a runaway process: once an insolation threshold is reached, activity is triggered (often by water), bringing volatiles (particularly CO$_{2}$) close to the surface and triggering more dust ejection. It is difficult to reduce the ejection rate to one that matches the observations by varying the various thermophysical parameters.

\begin{figure}
\resizebox{\hsize}{!}{\includegraphics[scale=0.5]{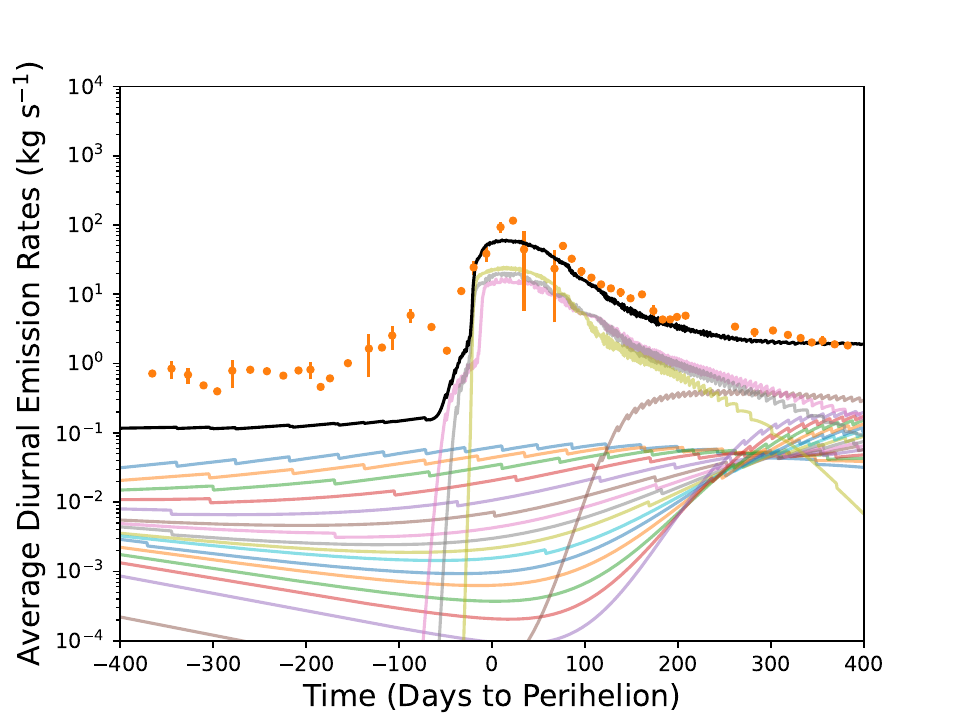}}%
\caption{Mean diurnal production rates of $CO_{2}$ for a model with $\delta=2$, f$_{CO_{2}}=0.1$, f$_{CO}=0.01$, $b=0.3D_{p}$, and a water-content dependent strength (see text for details), compared to the Rosetta data (solid points).}
\label{plot_ResultsQ_othervolatiles_cvariable_b0.3_strength-1}
\end{figure}

One parameter which we have not so-far discussed is material strength. A higher tensile strength would reduce the ejection rate by requiring a higher pressure, and therefore temperature, to be reached in the subsurface. However, too high a strength becomes impossible to reach and quenches the activity \citep{bischoff2023}. A more steeply decreasing strength with depth/size-scale than ours (which is based on the description of \citealp{Skorov.2012}) would help initiate activity by deep ejections, before slowing down its rate once CO$_{2}$-driven ejections take over close to the surface. A higher strength surface-layer could be evidence of a hard, sintered-ice or dense layer \citep{Spohn2015, Kossacki2015, Kofman2020}. However, our model shows multiple ejections per day at perihelion, so that this hardening process would have to be much faster than normally considered (e.g.~taking multiple orbital periods in the modelling of \citealp{Kossacki2015}). Alternatively, many laboratory experiments suggest that ice-rich material may have a higher intrinsic strength than dry dust even without sintering, due to a higher surface-energy for ice grains (e.g.~see \citealp{Tatsuuma2019}), or due to lower grain interconnectivity after the ice sublimates away \citep{Kreuzig.2024}. This could help reduce the frequency of CO$_{2}$-driven ejections, that mainly eject dust that still retains some water content, as compared to water-driven ejections, that eject dry dust. We tested this using a model run with a fixed tensile strength of 0.28 Pa (the average expected for $D_{p}=1$ cm-sized pebbles) for all layers containing water-ice, and 0.06 Pa (around five times less) for all layers not containing water-ice. This resulted in a very similar water-production curve to before, but the relatively-higher strength of water-rich material prevented it from being ejected by CO$_{2}$, leaving only water-driven ejections of dry dust. This reduced the ejection frequency (total dust mass-loss $\Delta$M$_{dust}=0.6$ of the observed) and the associated CO$_{2}$ outgassing (total mass-loss $\Delta$M$_{CO_{2}}=0.6$ of the observed), while maintaining enough CO$_{2}$ emission to match the observations very well at perihelion and afterwards, as shown in Fig.~\ref{plot_ResultsQ_othervolatiles_cvariable_b0.3_strength-1}. CO$_{2}$-emission is still too low before perihelion, as will be discussed in the next paragraph, but this ice-dependent strength model is very promising for reducing the role of CO$_{2}$-driven erosion whilst maintaining some ejections, and will be explored further in future work.

We now consider the emission rates at larger heliocentric distances. Although the mass fluxes are dominated by perihelion emission, the situation is also more thermophysically complex here, so that looking at the relatively simpler situation at lager heliocentric distances can help clarify the behaviour. Before perihelion, both CO$_{2}$ and CO show total outgassing rates that are low compared to the observations, and distributions that differ also. These differences are reduced in the model with low diffusivity, where CO$_{2}$ emission is more concentrated in the south and CO emission is more uniform, than in higher diffusivity models. Even for this case, however, total CO$_{2}$ emission is a bit low at the time of Rosetta's arrival, a year before perihelion, while the northern hemisphere seems relatively too active compared to the south. One possibility would be to reduce the northern CO$_{2}$ emission relative to the south (by burying the CO$_{2}$-front deeper with volatile-depleted fallback here, for example), but the overall low emission implies that we are actually missing some early southern hemisphere CO$_{2}$ emission. This is somewhat difficult to produce due to the fact that the southern hemisphere is experiencing polar night at this time, while thermal models such as here and in \citet{Groussin2024} suggest its interior to be very cold compared to the north (Fig.~\ref{plot_Results_sublimation fronts}, bottom right), both of which should limit outgassing. Nonetheless, low-latitude regions in the south (from the equator down to $\sim-45^{\circ}$) do receive limited amounts of insolation even early-on in the Rosetta period, and the situation could therefore be improved by additional early erosion in the south (from the top pebble eroding into small dust as discussed below, for example) bringing the CO$_{2}$ front closer to the surface. An alternative mechanism to supply early southern CO$_{2}$ emission is re-condensation. As pointed out in \citet{Groussin2024} and above, temperature inversions exist between the surface of the cold polar-nighttime southern hemisphere and its warmer interior, potentially allowing for CO$_{2}$ sublimating at depth to re-condense in shallower layers, ready to contribute to the early outgassing as soon as the Sun returns. Initial experiments resetting southern hemisphere CO$_{2}$ fractions to their starting levels at the second simulated aphelion passage indeed led to an earlier start to CO$_{2}$ outgassing, suggesting the validity of this mechanism. Re-condensed CO$_{2}$ might also help trigger ejections in the south slightly earlier, which would help bring the models more into line with the observations of large particle ejections within roughly $\pm100$ days of perihelion (see \citealp{Moreno2017}, Fig.~5), whereas ours mostly start at about $-50$ days. However, ejections are more sensitive to the diffusivity and strength than ice content. We plan to incorporate a proper treatment of re-condensation in a future version of the model.

\begin{figure}
\resizebox{\hsize}{!}{\includegraphics[scale=0.5]{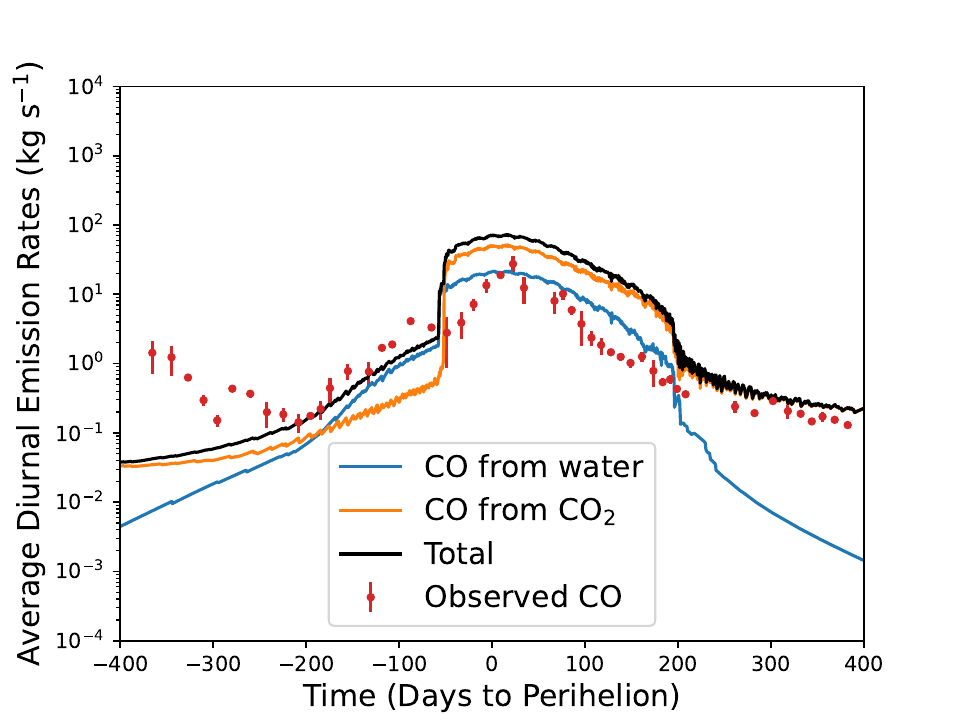}}%
\qquad
{\includegraphics[scale=0.5]{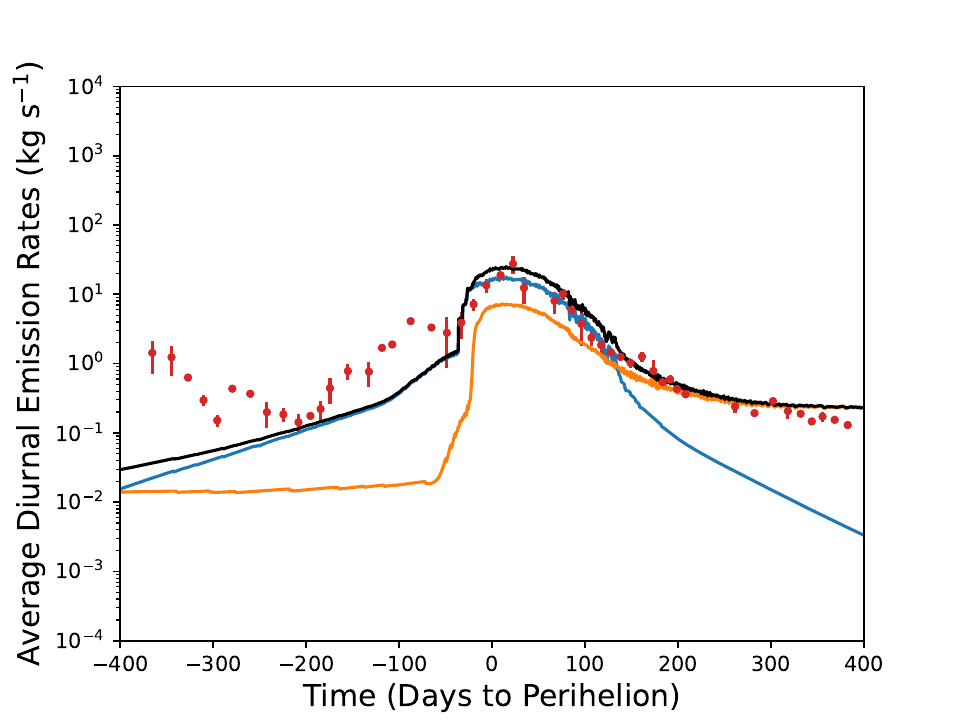}}%
\caption{Mean diurnal production rates for CO scaled from the production rates of water and CO$_{2}$ (see text for details) compared to the Rosetta data (solid points). Top: for the model with $\delta=2$, f$_{CO_{2}}=0.1$, f$_{CO}=0.01$, $b=0.1D_{p}$, and a fixed dust heat-capacity. Bottom: for a model with $\delta=2$, f$_{CO_{2}}=0.1$, f$_{CO}=0.01$, $b=0.3D_{p}$ and a water-content dependent strength.}
\label{plot_ResultsQ_CO_Rubin}
\end{figure}

Regarding the CO outgassing curves in particular, it should be noted that we model CO here as a separate ice-species, and this appears to work well after perihelion. However, CO may more likely be present trapped within other ices \citep{Gasc2017}, with \citet{Rubin2023} suggesting a partitioning  between water-ice and CO$_{2}$-ice with a ratio of around $70:30$. We test this briefly here by scaling the water and CO$_{2}$-outgassing curves by a total CO-to-water fraction of $4\%$ in this ratio. Note that this is not a full physical treatment of trapped ices or amorphous to crystalline phase-transitions in the numerical code (which will be considered in future work), but merely a scaling of the existing results. Figure \ref{plot_ResultsQ_CO_Rubin} shows the results for two different models: the $b=0.1D_{p}$, fixed dust heat-capacity run of figures \ref{plot_ResultsQ_cvariable_b0.1}, \ref{plot_ResultsQdust_cvariable_b0.1}, and \ref{plot_ResultsQ_othervolatiles_cvariable_b0.1}; and the water-content dependent strength model of figure \ref{plot_ResultsQ_othervolatiles_cvariable_b0.3_strength-1}. In both cases, the CO-production curve is well matched after perihelion, with the low CO$_{2}$-emission of the water-content dependent strength model also leading to a good peak outgassing rate at perihelion. Some pre-perihelion outgassing is missed in this model, however, due to the low activity around $-100$ days, also seen in Fig.~\ref{plot_ResultsQ_othervolatiles_cvariable_b0.3_strength-1}. This might be improved by the above-mentioned points on early southern CO$_{2}$-outgassing. By contrast, the model with our nominal strength law, a diffusivity of $b=0.1D_{p}$, and a high heat-capacity, does match the data well between around $-200$ and $-50$ days, but exceeds the measurements around perihelion, due to its very high activity here. Neither case is able to reproduce the initially high CO-outgassing that decreases with decreasing heliocentric distance prior to $-200$ d and that has an apparent origin in the northern hemisphere \citep{Hoang2019}, however, and this seems very difficult to model. One explanation could be that if very volatile material can survive in emitted chunks that fall-back onto the north \citep{Keller2017, Davidsson2022}, they could be reactivated when the north is illuminated at aphelion and decrease their activity towards perihelion as they exhaust their volatile supply. Attempts to model this scenario by adding numerical layers at the second aphelion containing ices at fractions of the original value have not, however, solved the issue. For small ice fractions, $\delta_{fallback}$, f$_{CO_{2} fallback}$ and f$_{CO fallback}$, CO sublimates quickly, before the next aphelion, and does not contribute any additional flux in the next orbit. When volatile fractions are increased, particularly f$_{CO_{2} fallback}$, it begins triggering ejections, quickly removing the entire fallback layer. This happens even for low assumed temperatures of the fallback material (down to 30 K) and low insolation levels, and is due to basal heating of the fallback by the heat retained in the northern hemisphere subsurface (see Fig.~\ref{plot_Results_sublimation fronts}, bottom left). Although it did not help to match the production curve, this phenomenon may be interesting in the context of the reactivation of fallback material: e.g. by contributing to the re-ejection and movement of chunks that retain some super volatiles as 'hopping' boulders \citep{Tang2025}. We note here that dust activity observed by Rosetta in the north implies at least some reactivation of fallback material. Generally, we do not model such reactivation explicitly, but the conclusions drawn above, about diffusivity and ice-content, and the implications for small dust activity below, also apply to fallback material.

Returning to the mysterious declining CO-outgassing early in the mission then, this may imply a complex mixing and desorption process \citep{Gasc2017}. Further modelling and laboratory work into volatile trapping and desorption is needed to explore this. Absent this, or additional measurements of 67P's CO-outgassing at large heliocentric distances that would help clarify the matter, we cannot definitively point to the location of CO-ice, but do note that the various models agree quite well with the available data at smaller heliocentric distances.

\subsection{Small dust}
\label{sec:discussion:mechanism}

Finally, one large problem remains with all the models studied here: activity is confined to southern latitudes around perihelion and only includes the ejection of particles of pebble-size or larger ($\gtrsim$ mm). This means we cannot generate the ubiquitous small-dust coma, seen by OSIRIS at all times and latitudes on 67P \citep{Marschall2020}. The relatively large depths of the water sublimation-front in the northern hemisphere ($10-30$ cm) are a consequence of this, and are also incompatible with its implied presence near the diurnal skin-depth around 1 cm or shallower \citep{Shi2016}. While the latter may be in-part solved by additional emission from water re-condensed near the surface during polar night or delivered by water-rich fallback, the former problem is more serious. In no simulations were we ever able to generate dust lifting in the northern hemisphere or far from perihelion. This actually agrees with the results of \citet{Xin2025}, who used a similar, ejecting dust-crust model. They only observed the onset of ejection activity inwards from 1.7 AU, and not in polar regions (for a modelled spherical nucleus with zero obliquity). This means that the small-dust ejecting mechanism must be solely responsible for erosion in the northern hemisphere. 

A number of possibilities exist for a mechanism that can eject small, sub-millimetre dust. In order to overcome the original cohesion bottleneck/activity paradox \citep{Kuehrt1994, Jewitt2019}, solutions must either increase the gas pressure, e.g.~inside low-diffusivity pebbles as in \citet{Fulle2019, Fulle2020}, or reduce the tensile strength. Examples of the latter include \citet{Zhang2024}, who suggest that the very low strength measured recently for meteorites \citep{Nagaashi2023} might facilitate ejections, while \citet{Kreuzig2025} and \citet{Schuckart2025} have recently proposed that individual dust grains, or aggregates thereof, become physically separated from the surface by the sublimation of small water-ice grains (or mantles) that were connecting them to their neighbours, essentially reducing the strength to zero. Both this and the \citet{Fulle2019} mechanisms produce very high dust fluxes and are therefore proposed to occur in spatially-limited, ice-rich patches (called water enriched bodies or WEBs by \citealp{Ciarniello2022, Ciarniello2023}), while the rest of the surface is relatively water-poor. This would not be compatible with the pebble and chunk ejection mechanism modelled here, which only works for high water-contents ($\delta\sim2$) and low CO$_{2}$-contents \citep{Attree2024b} everywhere. For spatially limited small-dust ejection to work in parallel with larger-scale chunk ejection then implies not two different ice-contents, but rather two different material structures (i.e.~differences in the microscopic location of the ice-content and arrangement of the pebble sub-structure). There would no longer be water enhanced- and water poor-bodies, but small-particle emitting, and dust-crust forming bodies. It is unknown what process would result in these two different populations of pebbles with different structures.

Alternatively, models that distribute small-dust activity over the whole cometary surface (such as in \citealp{Davidsson2022}) must find a way to generate a low and steady flux of this dust alongside the water (e.g.~with a flux that is smaller than the very fast erosion implied in WEBs). This could point to a dust-crust damping the water outgassing, combined with a statistical process of dust ejection whereby individual grains are occasionally broken off from inside the crust as gas flows through it. For any model, a dust mass-flux equal to the gas flux multiplied by the dust-to-ice ratio (which is compatible with the observed dust-mass fluxes for $\delta$ of a few: \citealp{Marschall2020, Moreno2017}) leads to the surface eroding at the same rate as the sublimation front retreats, meaning the two stay in constant balance. For our nominal model this retreat is about 4 to 7 cm per orbit over the northern hemisphere, as compared to the $~20-30$ cm-thick crust. The several metres worth of water-ice loss in the south are easily matched or exceeded by the ejection of chunks and pebbles.

Emphasising the point again then, the location and amount of erosion is of critical importance to the thermophysical models. A re-examination of OSIRIS coma images \citep{Marschall2020, Gerig2020} to test whether they are more compatible with the ejection of small dust by a homogenous process or a locally spotty one with a small emitting area would help constrain these processes.

\section{Conclusions}
\label{sec:conclusions}

Modelling the full, time-dependent emission rates of comet 67P/Churyumov-Gerasimenko with a one-dimensional thermophysical model remains challenging. The global production rates cannot be reproduced by any single patch on the surface, necessitating the consideration of multiple latitudes, as done here. Varying seasonal trends across cometary surfaces are thus shown to be very important in determining their outgassing and dust ejection patterns.

We have shown here that gas pressure in the subsurface can overcome material tensile strength, ejecting pebbles and chunks a few millimetres to around a decimetre in size, and resulting in water, CO$_{2}$, and CO outgassing rates that roughly match those observed by Rosetta. The ejecting dust-crust model naturally explains the results of NGA/T modelling that require water emission in the northern hemisphere to be relatively low and roughly constant or declining around perihelion, versus sharply increasing around perihelion in the south. This is due to volatile draining in the north leaving the sublimation fronts at large depths, while the south experiences 'blow-off' of the crust, leading to volatiles much closer to the surface during the active phase. The bulk of 67P's emissions (of all gas species and dust) then come from a relatively pristine southern hemisphere that undergoes strong erosion.

In order for this model to work, low gas diffusivities are needed (half-transmission thickness $b=0.3D_{p}$ particle radii at least, with a preference for  $b=0.1D_{p}$). We also find that a large heat-capacity and a steeply decreasing tensile strength with depth or ice-content are in best agreement with the outgassing data. However, the rates of ejection of dust and the more volatile gas species (CO$_{2}$ and CO) are still somewhat higher than the observations, even in these cases. In the south, it is difficult for models to achieve a balance between triggering activity and generating too much of it (particularly when CO$_{2}$ drives the ejections); while in the north, it remains challenging to generate activity at all, especially when fallback is considered.

Further, the model described here does not account for the ejection of dust smaller than the pebble size (i.e~sub-millimetre-sized dust), which requires a description of the substructure of the pebbles that is still relatively unknown. Nonetheless, by modelling pebble and chunk ejection here, we can place constraints on the processes that must drive the ejection of small dust. For example, if it is concentrated in a small surficial area of Water Enriched Bodies, then they must be resupplied in the north by fallback (i.e.~rather than being excavated by the ejection of chunks, which does not happen here). Further, as shown previously \citep{Attree2024b}, the non-WEB material cannot be completely water-depleted and CO$_{2}$-rich. Alternatively, if small dust is ejected from the whole cometary surface, then its ejection rate, and the associated water outgassing, must be small so as not to erode too much of the surface and contribute to too large dust- and water-fluxes.

Overall, we have improved our knowledge of and constraints upon the nature of the cometary activity mechanism, and highlight that the structure of the material, at both macro- and micro-scales, is critical in determining gas flow and pressure build-up.

\section*{Acknowledgements}

N.A. and P.G. acknowledge financial support from project PID2021-126365NB-C21 (MCI/AEI/FEDER, UE) and from the Severo Ochoa grant CEX2021-001131-S funded by MCI/AEI/10.13039/501100011033. D.B. and J.B. thank DFG for funding project BL 298/27-1. J.M. was financially supported by DFG project no. 517146316. This research was supported by the International Space Science Institute (ISSI) in Bern, through ISSI International Team project \#547 (Understanding the Activity of Comets Through 67P's Dynamics). The authors declare no conflicts of interest. We thank Imre Toth for a detailed and constructive review.

\section*{Data Availability}
The data underlying this article will be shared on reasonable request to the corresponding author.

%%%%%%%%%%%%%%%%%%%%%%%%%%%%%%%%%%%%%%%%%%%%%%%%%%



\bibliographystyle{mnras}
\bibliography{paper_references} % if your bibtex file is called example.bib

%%%%%%%%%%%%%%%%%%%%%%%%%%%%%%%%%%%%%%%%%%%%%%%%%%

% Don't change these lines
\bsp	% typesetting comment
\label{lastpage}
\end{document}